\documentclass[aps,pra,superscriptaddress,twocolumn,twocolumn,10pt]{revtex4-1}
\usepackage{amsmath}
\usepackage{latexsym}
\usepackage{graphicx}
\usepackage{amsfonts}
\usepackage{braket}
\usepackage{hyperref}
\usepackage{natbib}
\usepackage{amssymb}
\usepackage{nicefrac}
\usepackage{fancyhdr}
\usepackage[utf8]{inputenc}
\usepackage{dsfont}
\usepackage{colortbl}
\usepackage{xcolor}
\usepackage{subfigure}
\usepackage{psfrag}
\usepackage{tikz}
\usepackage{paralist}
\usepackage{nicefrac}
\usepackage{ulem}
\usepackage{placeins}
\usepackage{mathtools}
\usepackage{bbm}

\usetikzlibrary{arrows,shapes,decorations.pathmorphing}

\makeindex

\begin{document}

\title{Optimizing microwave photodetection: Input-Output theory}

\author{M. Sch\"ondorf}
\author{L. C. G. Govia}
\altaffiliation[Current Address: ]{Institute for Molecular Engineering, University of Chicago, Chicago, Illinois, USA}
\affiliation{Theoretical Physics, Saarland University, 66123
Saarbr{\"u}cken, Germany}
\author{M. G. Vavilov}
\author{R. McDermott}
\affiliation{Department of Physics, University of Wisconsin-Madison, Madison, WI 53706, USA}
\author{F. K. Wilhelm}
\affiliation{Theoretical Physics, Saarland University, 66123
Saarbr{\"u}cken, Germany}

\begin{abstract}
High fidelity microwave photon counting is an important tool for various areas from background radiation analysis in astronomy to the implementation of circuit QED architectures for the realization of a scalable quantum information processor. In this work we describe a microwave photon counter coupled to a semi-infinite transmission line. We employ input-output theory to examine a continuously driven transmission line as well as traveling photon wave packets. Using analytic and numerical methods, we calculate the conditions on the system parameters necessary to optimize measurement and achieve high detection efficiency. With this we can derive a general matching condition depending on the different system rates, under which the measurement process is optimal. 
\end{abstract}

\maketitle

\section{Introduction}
\label{sec:1}

Circuit quantum electrodynamics (cQED) has emerged as a powerful paradigm for the realization of quantum computational circuits in a scalable architecture \cite{clarke2008superconducting,blais2007quantum,chow2012universal,kelly2015state,brecht2015multilayer} as well as a demonstration of quantum radiation-matter interaction in the strong and ultra strong coupling regimes \cite{you2011atomic,hofheinz2009synthesizing,Niemczyk2010ultrastrong,baust2014ultrastrong}. Here, the lowest energy levels of a superconducting Josephson circuit play the role of an artificial atom, while thin film cavities and transmission lines are used to realize electromagnetic field modes. Strong coupling between the cavity fields and the artificial atom has been used to create strongly non classical states of the electromagnetic field \cite{hofheinz2009synthesizing,hofheinz2008generation,vlastakis2013deterministically,kirchmair2013observation,wang2016schrodinger,holland2015single}; in addition, coupling between these modes and the Josephson circuit can be used for high fidelity control \cite{PhysRevA.69.062320,krastanov2015universal} and measurement \cite{blais2004cavity,Sun2014photontracking,abdo2013directional,abdo2014josephson,riste2013deterministic,kindel2015generation,ribeill2011superconducting,hover2012superconducting,kinion2010microstrip,kinion2011superconducting}.

In conventional quantum optics at optical frequencies, detection of the electromagnetic mode is performed by a photon counter. The counter is typically modeled as an ensemble of two-level states that are weakly coupled to the light field \cite{Tannoudji}. Photon absorption is triggering a large, easily measured classical signal, and detector performance is expressed in terms of quantum efficiency and spurious dark count rate \cite{hadfield2009single}. In the microwave frequency range, conventional wisdom holds that there exists no material that can be photoionized by the lower frequency radiation. On the other hand, a variety of Josephson circuits are capable of detecting microwave photons down to the limit of a single photon with high efficiency \cite{narla2016robust,govia2012theory,chen2011microwave,poudel2012quantum,romero2009microwave,oelsner2016detection,peropadre2011approaching,fan2014nonabsorbing,inomata2016single,wong2015quantum,oelsner2016detection}.
Microwave photons can also be detected by lateral quantum dots\cite{wong2015quantum,kyriienko2016continuous}. In contrast to optical-frequency counters, Josephson-based microwave photon counters are realized as {\em single} effective two-level systems that couple strongly to the incident microwave field \cite{govia2012theory}. For this reason, they differ fundamentally from optical frequency counters. It is the purpose of this paper to explore the conditions for high-efficiency detection of propagating photons by these single, strongly coupled Josephson circuits. For the sake of completeness, we consider the Josephson photomultiplier (JPM), a current-biased junction capable of efficient detection of microwaves that are near resonant with the transition between the two lowest states in the metastable minima of the circuit potential. Previously, the JPM has been applied to investigation of temporal correlations of incident coherent and thermal microwave fields \cite{chen2011microwave}, and the JPM is currently under investigation for high fidelity measurement of single qubits \cite{govia2014high} and of multiqubit parity operators \cite{govia2015scalable}. Other approaches to single microwave photodetection include driven $\Lambda$ systems \cite{koshino2013implementation}. In this approach, the dressed states of a qubit-resonator system constitute an impedance-matched system, which absorbs an input photon with a near-unity efficiency \cite{koshino2015theory,koshino2016dressed}.

Here, we demonstrate that efficient microwave photon detection can be understood from a simple intuitive picture of rate matching, which has as its classical analog the usual impedance matching condition that provides for optimal power transfer in microwave circuits \cite{Pozar}. We present a general description of a transmission line directly coupled to a JPM, and explore the conditions that must be met to maximize detector quantum efficiency. Our results agree with those of \cite{romero2009photodetection}, where only a continuous drive input state was considered. Furthermore, our results extend beyond those of \cite{romero2009photodetection} as we include additional incoherent channels and study pulsed input states. A comparable condition was also found numerically in \cite{kyriienko2016continuous} for a different setting. Here they study a qutrit coupled to two transmission lines. In one of the transmission lines they induce a photon pulse. They show that the reflection coefficient is minimal if the coupling rate between the input transmission line and the qutrit is equal to the decay rate to the target state. These two rates can be translated into $\gamma_{\rm TL}$ and $\gamma_1$ in our description.

To describe our system, we use the input-output formalism \cite{gardiner1985input,clerk2010introduction}, a tool from the field of open quantum systems theory, that leads to generalized Heisenberg equations. The advantage of this approach is that it can be taken very far before specifying the form of the photon pulse in the transmission line making it versatile and its results broadly applicable. As a result, we can examine arbitrary states in the transmission line, including both continuous wave drive and wave packets with finite photon number. While equivalent to a density matrix approach, it is thus more effective for the problem at hand. 

The input-output formalism leads to a system of equations, from which we determine conditions on the system parameters that allow us to optimize detection efficiency. A sufficient set of these parameters can  be designed or even controlled in experiment such that this paper provides a guide towards practical implementation of the measurement of traveling photons using a JPM, achieving the optimal measurement efficiency experimentally possible.

This paper is organized as follows. In Sec. \ref{sec:2}, we present the system of interest and derive the corresponding equations of motion using input-output formalism. In Sec. \ref{sec:3}, we use a mean field approach that captures most of the quantum mechanical character of the system. We find the optimization conditions for continuous drive inputs, and for various pulsed waveforms. In Sec. \ref{sec:4}, we solve the equations by substituting operators with their corresponding expectation values. This simplification leads to rate equations, the solution of which yields a general matching condition for measurement optimization, which agrees with the result of Sec. \ref{sec:3}. In Sec. \ref{sec:5}, we present our conclusions.

\section{System and Equations of Motion}
\label{sec:2}

\begin{figure}
\includegraphics[width=0.5\textwidth]{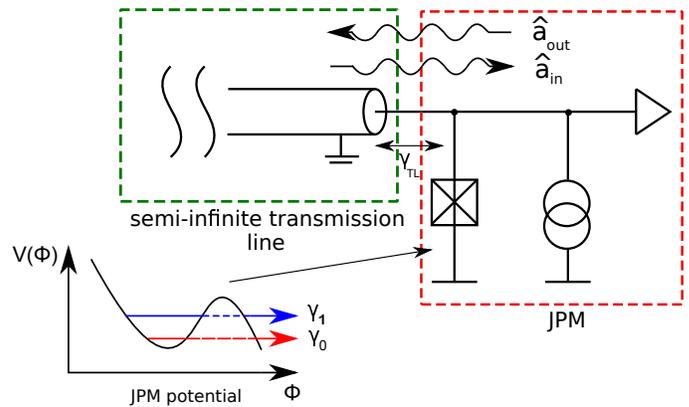}
\caption{\label{fig:JPM_TL} System schematic. The JPM is directly coupled to a transmission line which excites the JPM by an incoming photon flux. The potential of the JPM is a tilted washboard with two quasi-bound states in the local minima.}
\end{figure}

The system of interest is a microwave transmission line directly coupled to a JPM. The system Hamiltonian is written as
\begin{align}
 \hat H = \hat H_{\rm JPM} + \hat H_{\rm TL} + \hat H_{\rm INT},
\end{align}
where $\hat H_{\rm JPM}$ denotes the Hamiltonian of the JPM, $\hat H_{\rm TL}$ is the bare transmission line Hamiltonian, and $\hat H_{\rm INT}$ describes the interaction between the transmission line and the JPM. The JPM is realized through a current biased Josephson junction and is described by a tilted washboard potential \cite{martinis1985energy}, from which one can isolate two quasi-bound energy levels $\ket{0}$ and $\ket{1}$, with associated Hamiltonian
\begin{align}
 \hat H_{\rm JPM} = -\hbar \omega_0 \frac{\hat \sigma_z}{2}.
\end{align}
Here, $\omega_0$ is the transition frequency and $\hat \sigma_z = \left[\hat \sigma^{-},\hat \sigma^{+}\right]$ is the usual Pauli-Z operator with
\begin{align}
\hat \sigma^{-} = \ket{0}\bra{1} \hspace{0.5cm} \hat \sigma^{+} = \ket{1}\bra{0}.
\end{align}
Note that the local minima in the JPM potential are physically equivalent and only transitions between them can be detected \cite{likharev1985theory} (see Fig. \ref{fig:JPM_TL}). Both states can tunnel to the continuum with rate $\gamma_0$ and $\gamma_1$, respectively. For our description, we represent the continuum by a fictitious measurement state $\ket{m}$. Incoherent tunneling to the $\ket{m}$ state corresponds to generation of a measurable voltage pulse. Absorption of a resonant photon induces a transition from $\ket{0}$ to $\ket{1}$, which tunnels rapidly to the continuum since $\gamma_1 \gg \gamma_0$; this system can thus be used to count incoming photons. 

Quantization of the transmission line \cite{blais2004cavity} leads to the usual multimode harmonic oscillator Hamiltonian
\begin{align}
\hat H_{\rm TL} = \hbar \int_{0}^{\infty}   |f(\omega)|^2 \omega \hat a^{\dag}(\omega)\hat a(\omega) {\rm d}\omega.
\end{align}
Here, $\omega$ is the frequency of the transmission line mode and $\hat a^{\dag}(\omega)$, $\hat a(\omega)$ are the bosonic creation and annihilation operators for a photon at frequency $\omega$, respectively. 
$f(\omega)$ is the envelope of the incoming radiation in frequency space and has units $1/\sqrt{\omega}$ which in our case is assumed to be real (for more detail on how to model incoming radiation fields in the Heisenberg picture see \cite{baragiola2012n} and \cite{divincenzo2013multi}).  

The interaction between the JPM and the transmission line arises from the additional bias on the JPM caused by the transmission line current (see Fig. \ref{fig:JPM_TL}). This leads to a dipole interaction between the JPM states and the transmission line described by the Hamiltonian
\begin{align}
\hat H_{\rm INT} = \Delta \hat I \frac{\Phi_0}{2\pi} \hat \varphi_J,
\label{interaction}
\end{align}
where $\Phi_0 \equiv h/2e$ is the magnetic flux quantum and $ \Delta \hat I$ and $\hat \varphi_J$ describe the additional quantized current coming from the transmission line and the quantized phase of the JPM, respectively. To derive expressions for $\Delta \hat I$ and $\hat \varphi_J$ we use standard circuit quantization, which yields \cite{geller2007quantum,johansson2010dynamical}
\begin{align}
\Delta \hat I &= \sqrt{\frac{\hbar \omega_s}{4\pi Z_0}} \int_{0}^{\infty} {\rm d}\omega f(\omega)\left(\hat a^{\dagger}(\omega)+\hat a(\omega)\right)\label{current}\\
\hat \varphi_J &= \frac{i}{\sqrt{2}}\left(\frac{2E_C}{E_J}\right)^{\frac{1}{4}} \left(\hat \sigma^{+}-\hat \sigma^{-}\right).
\label{phi}
\end{align}
Here $Z_0$ is the transmission line impedance at the characteristic frequency $\omega_s$ of the incoming signal; $E_C = (2e)^2/2C_J$ is the Cooper pair charging energy, with the junction self-capacitance $C_J$; and $E_J = \hbar I_c/2e$ is the Josephson coupling energy, where $I_c$ is the critical current of the junction. Inserting expressions \eqref{current} and \eqref{phi} into \eqref{interaction}, we obtain the quantized interaction Hamiltonian
\begin{align}
\hat H_{\rm INT} = i \hbar \sqrt{\frac{\gamma_{\rm TL}}{2\pi}} \int_{-\infty}^{\infty} {\rm d}\omega f(\omega) \left[\hat a^{\dag}(\omega) \hat \sigma^{-} - \hat \sigma^{+} \hat a(\omega)\right],
\label{interaction_hamilton}
\end{align} 
where $\gamma_{\rm TL} =  \omega_s Z_J/4 Z_0$ describes the coupling rate between the transmission line and the JPM. The expression for $\gamma_{\rm TL}$ includes the junction impedance $Z_J = 1/\omega_s C_J$.

For this derivation (see Appendix \ref{app:1}) we applied the rotating-wave-approximation (RWA) \cite{WallsMilburn}, which leads to a continuous Jaynes-Cummings interaction \cite{jaynes1963comparison} and allows us to put the lower limit of integration to $-\infty$ instead of $0$. We further assumed that the coupling is constant over all modes, which is the first Markov approximation \cite{QuantumNoise}. Since the interaction is described by \eqref{interaction_hamilton}, we can use standard input-output formalism \cite{gardiner1985input} to derive the quantum mechanical Langevin equation for an arbitrary JPM operator $\hat S$ (see after eq. \eqref{Langevin_Lindblad} for further remarks)
\begin{align}
\begin{split}
\dot{\hat S}(t) &= \frac{i}{\hbar}\left[\hat H_{\rm JPM}, \hat S(t)\right]\\ &- \left[\hat S(t), \hat \sigma^{+}(t)\right]\left\{ \frac{\gamma_{\rm TL}}{2}\hat \sigma^{-}(t) - \sqrt{\gamma_{\rm TL}} \hat a_{\rm in}(t)\right\} \\ &+ \left\{\frac{\gamma_{\rm TL}}{2}\hat \sigma^{+}(t) - \sqrt{\gamma_{\rm TL}} \hat a_{\rm in}^{\dag}(t)\right\}\left[\hat S(t), \hat \sigma^{-}(t) \right],
\end{split}
\label{Langevin}
\end{align}
with input field operator defined as
\begin{align}
\hat a_{\rm in}(t) \equiv -\frac{i}{\sqrt{2\pi}} \int_{-\infty}^{\infty} {\rm d}\omega \exp \left[-i \omega \left(t-t_0\right)\right] f(\omega) \hat a_{t_0}(\omega),
\label{a_in_def}
\end{align}
where $\hat a_{t_0}(\omega)$ is the field operator at time $t=t_0$ and $f(\omega)$ is again the envelop of the incoming radiation. Without loss of generality, we set the starting point of the interaction to zero, $t_0 = 0$. Our system satisfies the standard input-output relation \cite{WallsMilburn}
\begin{align}
\hat a_{\rm out}(t) + \hat a_{\rm in}(t) = \sqrt{\gamma_{\rm TL}}\hat \sigma^{-}(t),
\label{inout_Relation}
\end{align}
where the output field operator is defined as
\begin{align}
\hat a_{\rm out}(t) = \frac{i}{\sqrt{2\pi}} \int_{-\infty}^{\infty} {\rm d}\omega \exp\left[- i \left(t-t_1\right)\right] f(\omega) \hat a_{t_1}(\omega).
\end{align}
Here, $\hat a_{t_1}(\omega)$ is similar to $\hat a_{t_0}(\omega)$ in that it is defined as the field operator at a time $t_1>t_0$ after the interaction between transmission line and JPM is turned on. Here $f(\omega)$ describes the envelop of the outgoing radiation.

Up to now we have not considered incoherent decay channels of the JPM. We include them using the standard Lindblad formalism. The Lindblad operator that describes tunneling from the excited state to the continuum (measurement process) is
\begin{align}
\hat L_1 = \sqrt{\gamma_1} \ket{m}\bra{1},
\end{align}
with tunneling rate $\gamma_1$, where the state $\ket{m}$ represents all states outside the potential well of the quasi-bound states. Another incoherent channel is given by dark counts
\begin{align}
\hat L_0 = \sqrt{\gamma_0} \ket{m} \bra{0},
\end{align}
a tunneling with rate $\gamma_0$ from the ground state of the JPM into the measurement state. We also take into account the possibility of relaxation from $\ket{1}$ to $\ket{0}$ through energy loss to the environment. This process is described by the Lindblad operator
\begin{align}
\hat L_{\rm rel} = \sqrt{\gamma_{\rm rel}} \ket{0}\bra{1},
\end{align}
where $\gamma_{\rm rel}$ is the relaxation rate. This rate only includes emission into the intrinsic environment of the JPM, since emission back to the transmission line is already built into the input-output equations. Finally, we assume that the JPM has the possibility to reset after a measurement, such that multiple measurements are possible. The reset is described by the operator
\begin{align}
\hat L_{\rm res} = \sqrt{\gamma_{\rm res}} \ket{0}\bra{m},
\end{align}
 where $\gamma_{\rm res}$ is the reset rate. The reset process brings the JPM from the measurement state $\ket{m}$ back to the ground state $\ket{0}$.

To include these Lindblad channels in the above Langevin equation, we use the adjoint master equation \cite{Breuer}
\begin{align}\label{Lindblad}
\begin{split}
\dot{\hat S}(t) &= \frac{i}{\hbar} \left[H_{\rm JPM}, \hat S(t)\right] \\ &\hspace{-0.5cm} + \sum_k\left(\hat L_k^{\dag}S(t)\hat L_k - \frac{1}{2} \hat S(t) \hat 
L_k^{\dag}\hat L_k - \frac{1}{2} \hat L_k^{\dag}\hat L_k \hat S(t)\right), 
\end{split}
\end{align}
with $k \in \{0,1,{\rm rel},{\rm res}\}$ and $\hat S$ an arbitrary JPM operator. Combining \eqref{Langevin} and \eqref{Lindblad}, we obtain a Langevin-Lindblad master equation that describes the coherent and incoherent dynamics of an arbitrary system operator
\begin{align}\label{Langevin_Lindblad}
\begin{split}
\dot{\hat S}(t) &= \frac{i}{\hbar}\left[\hat H_{\rm JPM}, \hat S(t)\right] \\ &\hspace{-0.4cm}- \left[\hat S(t), \hat \sigma^{+}(t)\right]\left\{ \frac{\gamma_{\rm TL}}{2}\hat \sigma^{-}(t) - \sqrt{\gamma_{\rm TL}} \hat a_{\rm in}(t)\right\}\\ &\hspace{-0.4cm}+ \left\{\frac{\gamma_{\rm TL}}{2}\hat \sigma^{+}(t) - \sqrt{\gamma_{\rm TL}} \hat a_{\rm in}^{\dag}(t)\right\}\left[\hat S(t), \hat \sigma^{-}(t) \right]\\  &\hspace{-0.4cm}+ \sum_k \left(\hat L_k^{\dag}S(t)\hat L_k  - \frac{1}{2}\left[\hat S(t) \hat 
L_k^{\dag}\hat L_k + \hat L_k^{\dag}\hat L_k \hat S(t) \right] \right).
\end{split}
\end{align}
All of the above Lindblad operators describe loss channels of the JPM. Note that this is written for as an equation for JPM operators, hence the transmission line operators act as noise sources like in the classical Langevin equation. They are operator-valued to reflect the quantum nature of the noise (for more details see \cite{QuantumNoise}). In general, the transmission line can also evolve incoherently; however, the rates for these processes are slow compared to JPM processes \cite{goppl2008coplanar},\cite{megrant2012planar}, so they are ignored in our calculations.

We are interested in the occupation probabilities of the different JPM states, defined by the projection operators
\begin{align}
\begin{split}
\mathcal{\hat P}_0 \equiv \ket{0}\bra{0}\hspace{0.5cm} 
\mathcal{\hat P}_1 \equiv \ket{1}\bra{1}\hspace{0.5cm}
\mathcal{\hat P}_m \equiv \ket{m}\bra{m}.
\end{split}
\end{align}
To obtain a complete system of equations, we must also include the system raising and lowering operators $\hat \sigma^{-}$, $\hat \sigma^{+}$. Putting these five operators into equation \eqref{Langevin_Lindblad} leads to a set of coupled ordinary differential equations
\begin{subequations}
\begin{align}
\label{System1} 
\dot{\hat \sigma}^{-} &= -i\omega_0 \hat \sigma^{-}+\sqrt{\gamma_{\rm TL}}\hat \sigma_z \hat{a}_{\rm in} - \frac{\tilde \gamma}{2} \hat \sigma^{-} \\
\label{System2}
\dot{\hat \sigma}^{+} &= i\omega_0 \hat \sigma^{+} +\sqrt{\gamma_{\rm TL}}\hat{a}_{\rm in}^{\dagger} \hat \sigma_z - \frac{\tilde \gamma}{2} \hat \sigma^{+}   \\
\label{System3}
\dot{\hat{\mathcal P}}_0 &= -\gamma_0  \mathcal{\hat P}_0 + (\gamma_{\rm TL}+\gamma_{\rm rel}) \mathcal{\hat P}_1 -\sqrt{\gamma_{\rm TL}} \mathcal{\hat W}+\gamma_{\rm res} \mathcal{\hat P}_m   \\
\label{System4}
\dot{\hat{\mathcal P}}_1 &= -(\gamma_{\rm TL}+\gamma_{\rm rel}+\gamma_1) \mathcal{\hat P}_1  +\sqrt{\gamma_{\rm TL}} \mathcal{\hat W}   \\
\label{System5}
 \dot{\hat{\mathcal P}}_m &= \gamma_0 \mathcal{\hat P}_0+\gamma_1 \mathcal{\hat P}_1 -\gamma_{\rm res} \mathcal{\hat P}_m, 
\end{align}
\end{subequations}
where $\tilde \gamma$ is defined as $\tilde \gamma \equiv \gamma_{\rm TL}+\gamma_0+\gamma_1+\gamma_{\rm rel}$ and $\mathcal{\hat W} \equiv \hat{a}_{\rm in}^{\dagger}\hat\sigma^{-} + \hat \sigma^{+} \hat{a}_{\rm in}$. All operators are time-dependent, since we are in the Heisenberg picture. Here and in the following, however, we will only indicate this time dependence explicitly when it is necessary for clarity.

It should be noted that up to this point we have made no assumptions about the input field $\hat a_{\rm in}$, such that the derived system of equations describes a completely general pulse/drive. This allows us to examine different incoming fields in the transmission line, including both continuous drive and various forms of pulses.

\section{Mean Field Approach}
\label{sec:3}

In this section, we use a mean field approach (see \cite{kocabacs2012resonance}) to simplify equations \eqref{System1}-\eqref{System5}. This approach includes first order correlations between the transmission line and the JPM. It is based on the assumption that the transmission line stays in a coherent state described by a single amplitude $\alpha$. It tacitly assumes that not only $\hat{a}|\alpha\rangle=\alpha|\alpha\rangle$ as usual but also 
$\hat{a}^\dagger|\alpha\rangle=\alpha^\ast|\alpha\rangle$ or, alternatively, $\left\langle \alpha^\prime |\alpha \right\rangle= 0$ for $\alpha' \neq \alpha$, thus assuming a large initial coherent state with $|\alpha|\gg 1$ (see \cite{kocabacs2012resonance}).

 An important point for the whole section is that the variable $|\alpha|^2$ in our case is the amplitude of a photon flux whereas in the standard case it denotes the actual photon number. This fact arises from the usual formalism used in input-output theory, where field creation and anhilation operators are not unitless (see e.g.\cite{WallsMilburn}). The actual photon number that hits the detector during the measurement time interval $t_m$ is then given by $n = |\alpha|^2 \omega_0 t_m$ (see App. \ref{app:photon_flux}). Hence the condition for the validity of the approximation in our case reads $|\alpha|^2 \omega_s t_m \gg 1$. Note that some of the results we show in the following extrapolate to regimes where this condition is not fulfilled, e.g. we start with $|\alpha|^2 =0 $ in some plots, but the key results are in the regime where the approximation holds.

In the following, we only consider one measurement event ($\gamma_{\rm res} = 0$) and look at the measurement probability to define the efficiency of the counter, since this value corresponds to the efficiency in the multi-count case (for short enough reset time). Additionally, we neglect dark counts ($\gamma_0 = 0$) since the typical dark count rates of a JPM do not change the results significantly, as we will see in Sec \ref{sec:4}. For simplicity we also assume that we do not have any relaxation ($\gamma_{\rm rel} = 0$).

We are especially interested in the choice of $\gamma_{\rm TL}$, that maximizes the measurement probability. We refer to this rate as $\gamma_{\rm TL}^{\rm max}$.
\begin{figure*}[t!]
\includegraphics[width=0.49\textwidth]{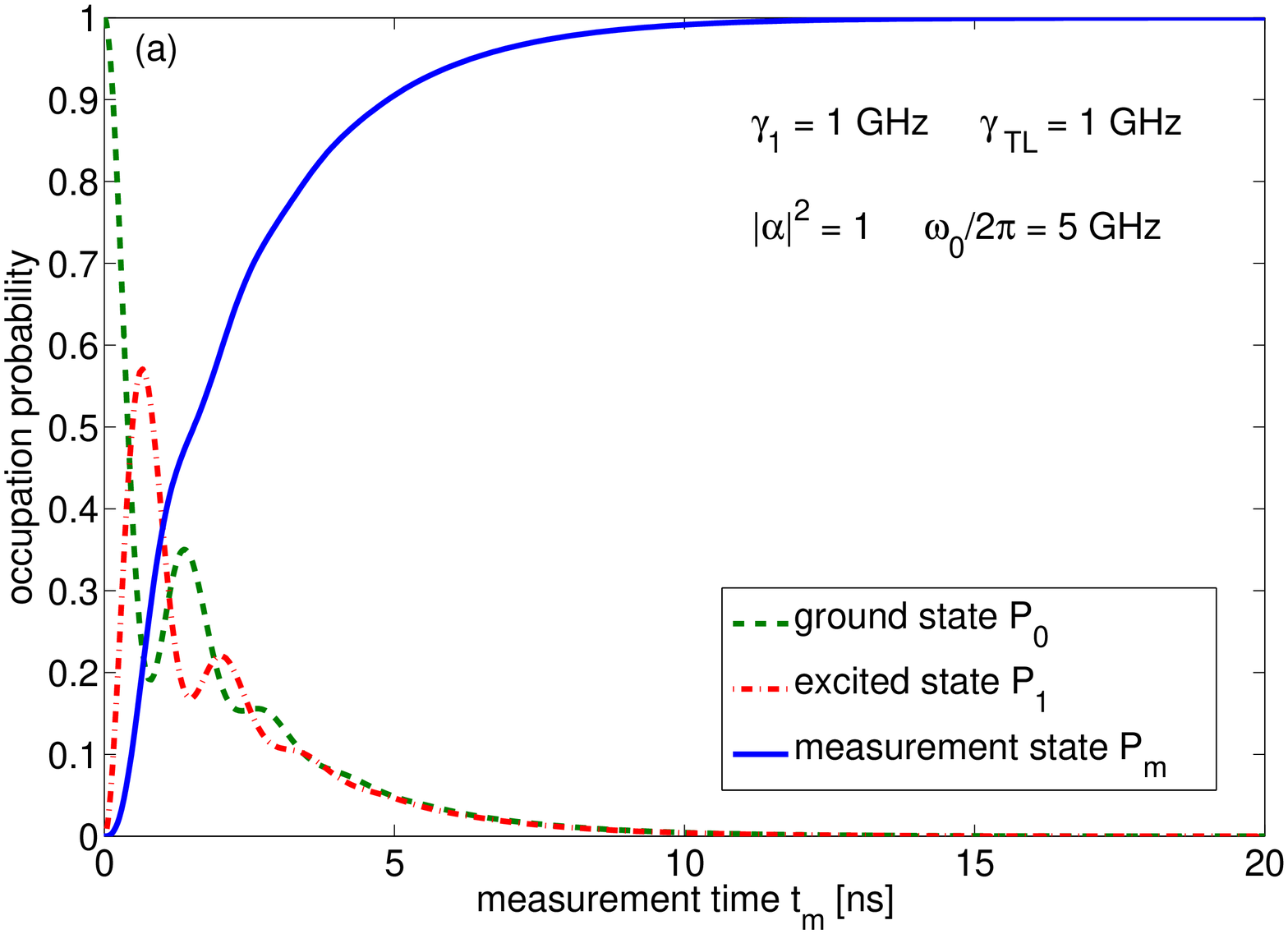}
\includegraphics[width=0.49\textwidth]{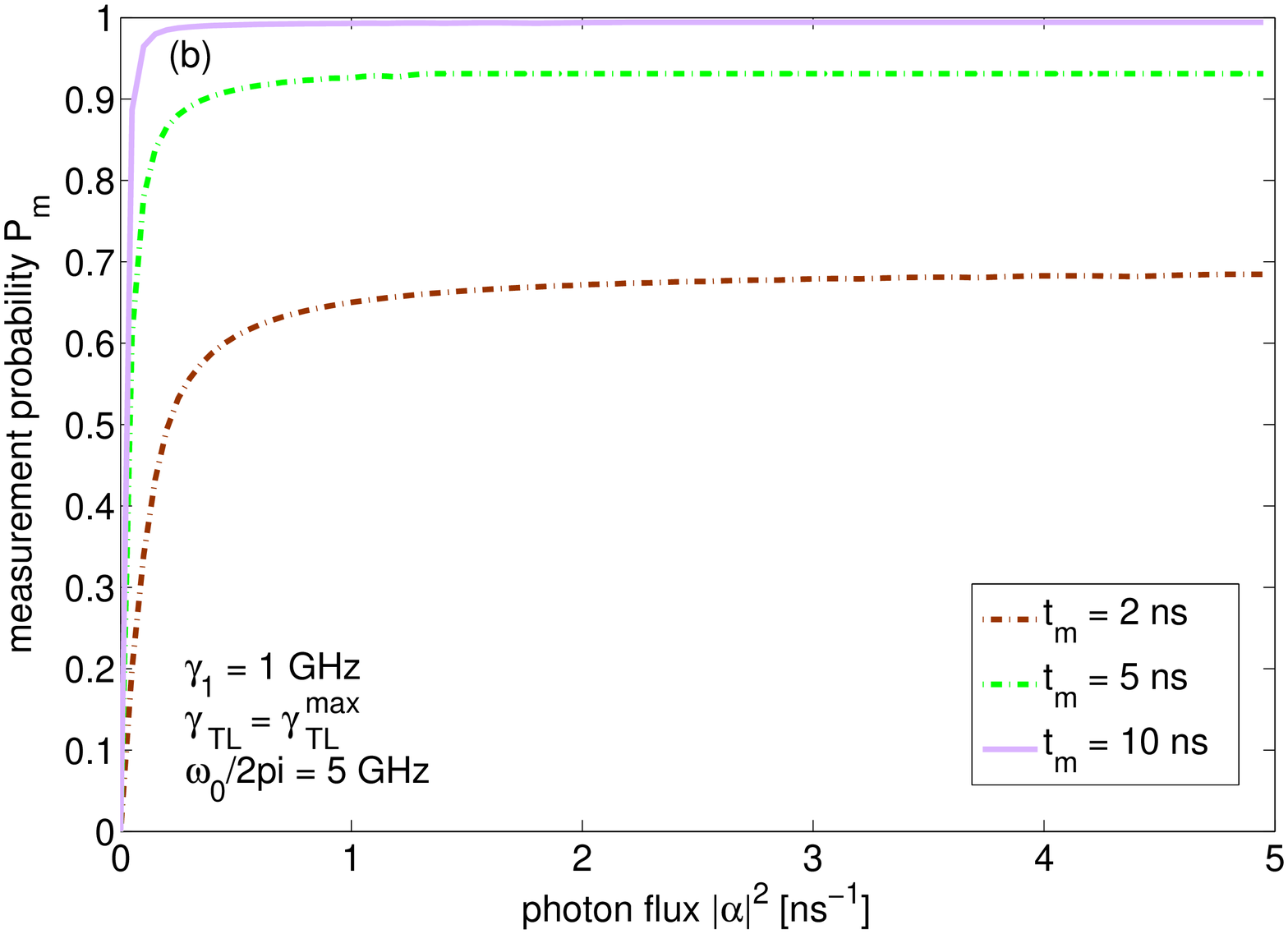}
\includegraphics[width=0.49\textwidth]{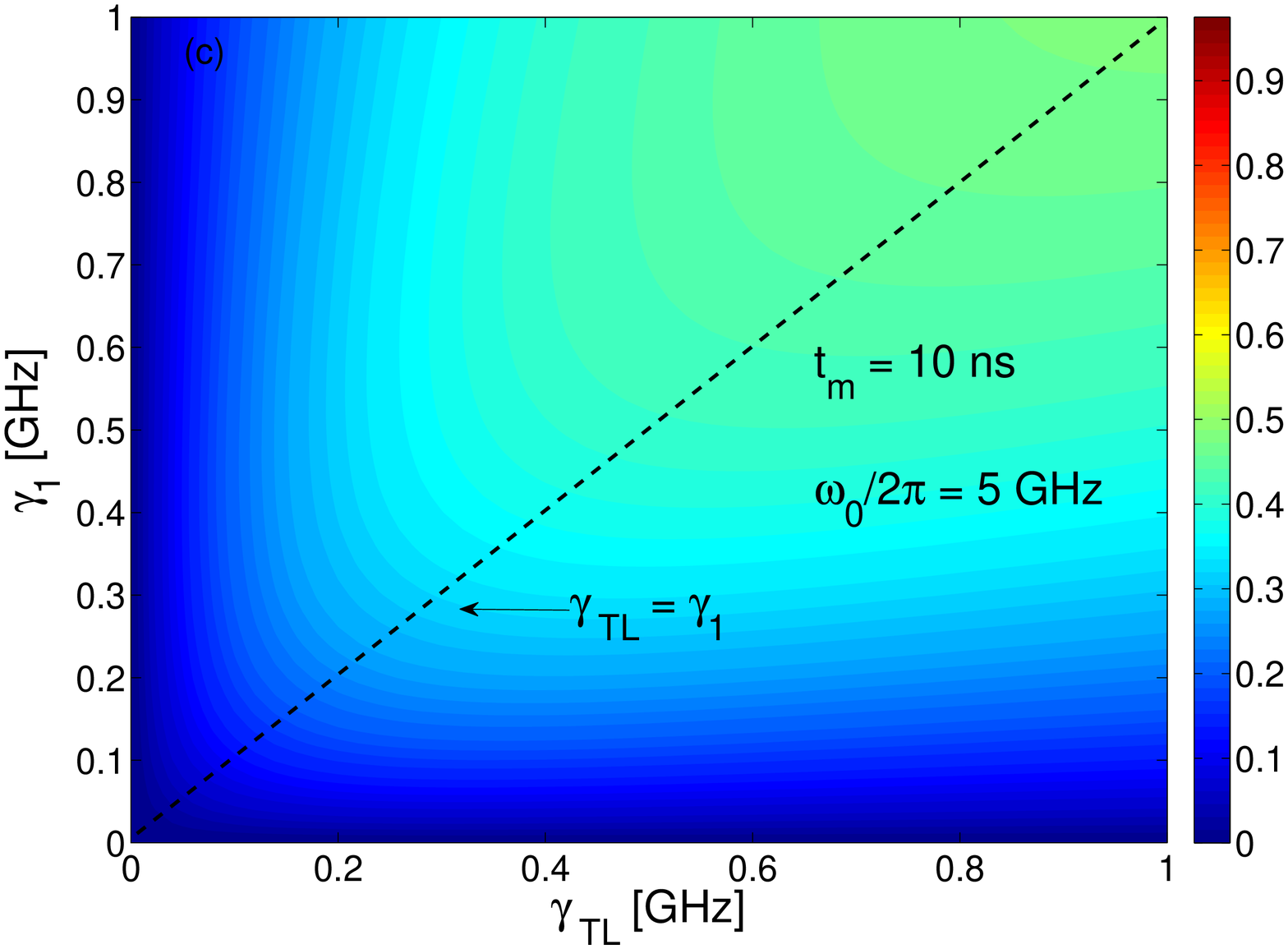}
\includegraphics[width=0.49\textwidth]{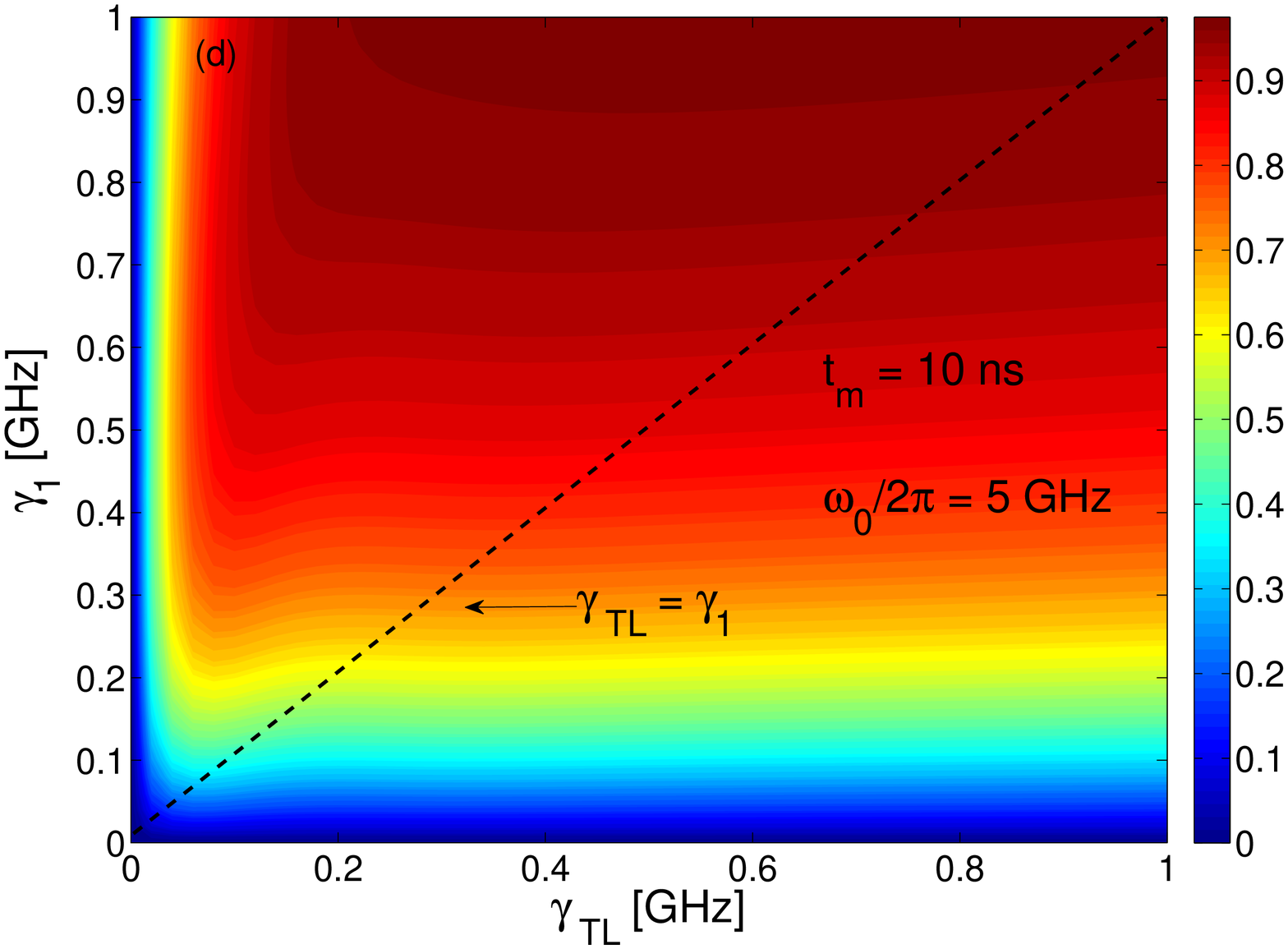}
\caption{\label{fig:continuous} (a) Occupation probabilities as a function of the measurement time $t_m$. (b) Measurement probability as a function of the rate of incoming photons for optimal rate choice $\gamma_{\rm TL} = \gamma_{\rm TL}^{\rm max}$ (before steady state is reached). One sees a saturation at around when the rate of incoming photons exceeds the measurement rate, such that increasing the rate of incoming photons does not further increase the measurement probability. (c),(d) Measurement probability versus the rates $\gamma_{\rm TL}$ and $\gamma_1$ after $t_m = 10$ ns (before stationary state is reached) for two different values of $|\alpha|^2$. (c) For small values of $|\alpha|^2$ (0.5 photons during $t_m$), the optimal measurement regime coincides with the matching condition \eqref{matching_simple} found in Section II. (d) For high values of $|\alpha|^2$ (50 photons during $t_m$), we see a plateau behavior, such that the measurement probability is independent of $\gamma_{\rm TL}$.}
\end{figure*}
\subsection{Continuous Input}
\label{sec:4A}

We assume that we have a continuous, coherent drive at frequency $\omega_0$ (such that the signal frequency $\omega_s$ is equal to $\omega_0$) and photon flux amplitude $\alpha$, such that the initial state reads
\begin{align}
\ket{\Phi (t=0)} = \ket{0}_{\rm JPM} \otimes \ket{\alpha}_{\rm TL} = \ket{0,\alpha},
\label{initial}
\end{align}
where the JPM is arranged in the ground state before measurement and the transmission line is in a coherent state of amplitude $\alpha$ and frequency $\omega_0$. We can take the expectation value in the system of equations \eqref{System1}-\eqref{System5} with respect to state \eqref{initial} (note that the time dependence is included in the operators, such that $\ket{\Phi}$ stays constant). To trace out the transmission line degrees of freedom, we apply $\hat a_{\rm in}$ to the right and $\hat a^{\dag}_{\rm in}$ to the left, which gives
\begin{align}
\begin{split}
\hat a_{\rm in} \ket{0,\alpha_{\omega_0}} &= - \frac{i}{\sqrt{2\pi}} \int_{-\infty}^{\infty}{\rm d}\omega  \exp[-i\omega t] f(\omega)\hat a(\omega) \ket{0,\alpha} \\
&= -\frac{i}{\sqrt{2\pi}} \alpha \sqrt{\omega_0} \exp\left[-i\omega_0 t\right] \ket{0,\alpha}
\end{split},
\end{align}
since a single mode drive is described by a $\delta$-function in frequency space for a continuous drive at frequency $\omega_0$: $f(\omega) = \sqrt{\omega_0}\delta(\omega-\omega_0)$.
In addition we apply the transformation $\hat \sigma^{-} \longmapsto \exp[-i\omega_0 t]\hat\sigma^{-}$ and $\hat \sigma^{+} \longmapsto \exp[i\omega_0 t]\hat \sigma^{+}$ in order to make the equations time independent. After these steps, we finally end up with equations of motion for the expectation values of the JPM operators:
\begin{figure}
\includegraphics[width=0.49\textwidth]{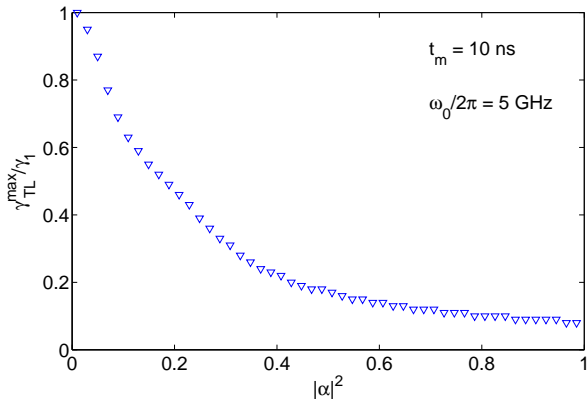}
\caption{\label{fig:onemode_gammaTL} Dependence of $\gamma_{\rm TL}^{\rm max}/\gamma_1$ on $|\alpha|^2$ for the continuous drive case. For small values of $|\alpha|^2$, the optimal regime is the matching condition $\gamma_{\rm TL} = \gamma_1$ we will also find analytically in Sec. \ref{sec:4} (see Eq. \eqref{matching_simple}). For higher values of $|\alpha|^2$ the optimal measurement regime shifts to smaller ratios $\gamma_{\rm TL}^{\rm max}/\gamma_1$, since the Rabi frequency is proportional to $\sqrt{\gamma_{\rm TL}|\alpha|^2}$.}
\end{figure}
\begin{subequations}
\begin{align}
\left<\dot{\hat \sigma}^{-}\right> &= - \frac{\tilde \gamma}{2} \left<\hat \sigma^{-}\right>  - \rm{i} \frac{\omega_R}{2}  \left(\left<\mathcal{\hat P}_0\right> - \left<\mathcal{\hat P}_1\right>\right) \label{continuous_1} \\ 
\left<\dot{\hat \sigma}^{+}\right> &= - \frac{\tilde \gamma}{2} \left<\hat \sigma^{+}\right>+ \rm{i} \frac{\omega_R}{2}  \left(\left<\mathcal{\hat P}_0\right> - \left<\mathcal{\hat P}_1\right>\right)\label{continuous_2} \\   
\left<\dot{\hat{\mathcal P}}_0\right> &= \gamma_{\rm TL} \left<\mathcal{\hat P}_1\right> - \rm{i} \frac{\omega_R}{2} \left(\left<\hat \sigma^{-}\right>  - \left<\hat \sigma^{+}\right> \right)\label{continuous_3}\\ 
\left<\dot{\hat{\mathcal P}}_1\right> &= -\tilde \gamma \left<\mathcal{\hat P}_1\right> + \rm{i} \frac{\omega_R}{2} \left(\left<\hat \sigma^{-}\right>  - \left<\hat \sigma^{+}\right> \right) \label{continuous_4}\\   
\left<\dot{\hat{\mathcal P}}_m\right> &= \gamma_1 \left<\mathcal{\hat P}_1\right>\label{continuous_5},
\end{align}
\end{subequations}
where $\omega_R \equiv \sqrt{2|\alpha|^2\gamma_{\rm TL}\omega_0/\pi}$ denotes the Rabi frequency, and where we have used the relation $\left<\hat \sigma_z\right> = \left<\mathcal{\hat P}_0\right>-\left<\mathcal{\hat P}_1\right>$ to eliminate $\left<\hat \sigma_z\right>$. This system of equations can be solved numerically (see Fig. \ref{fig:continuous}).

We are mostly interested in the measurement probability $\left<\mathcal{\hat P}_m\right>$. For every choice of parameters, the measurement probability reaches unity after some time since we assume a continuous drive (see Fig. \ref{fig:continuous}(a)), so that energy transfer to the JPM continues for as long as needed to tunnel to the measurement state. The switching time depends on the choice of parameters, and we see that for small values of $|\alpha|^2$, the condition that minimizes this time is $\gamma_{\rm TL}=\gamma_1$ which we refer to as the matching condition. In the next section we will see, that we find the same matching condition analytically with a less rigorous approximation (see Fig.\ref{fig:continuous}(c) and Fig. \ref{fig:onemode_gammaTL}). 
For higher values of $|\alpha|^2$, the matching condition shifts to smaller values of $\gamma_{\rm TL}$ (see Fig. \ref{fig:onemode_gammaTL}). If $\omega_r$ $\gg$ $\gamma_1$, the system dynamics are much faster than the measurement process, such that the JPM likely oscillates back to the ground state before tunneling from the excited state to the measurement state. On the other hand, if $\gamma_1$ $\gg$ $\omega_r$, measurement can be seen as a continuous projection and therefore freezes the system dynamics. This effect is well known as the quantum zeno effect \cite{misra1977zeno,itano1990quantum,damborenea2002measurement,jacobs2007continuous,rossi2008quantum,wang2008quantum,helmer2009quantum}.
If we match the rates and look at the correlation between measurement time and the rate of incoming photons, we see a saturation at the point when the rate of incoming photons becomes greater than the measurement time, since then the arrival of a photon at the detector during the measurement time is guaranteed (see Fig. \ref{fig:continuous}(b)). This means that adding more photons per time interval does not increase the measurement probability, since the JPM can only measure one photon (see Fig. \ref{fig:continuous}(b)). The measurement probability is one at this saturation point if the measurement time is longer than the required time for a tunneling process, and smaller than one otherwise (see Fig. \ref{fig:continuous}(b)).
Moreover, we find that for high values of $|\alpha|^2$ there is a large region where the measurement probability is independent of $\gamma_{\rm TL}$ and only varies with $\gamma_1$ (see Fig. \ref{fig:continuous}(d)), corresponds to the classical regime. Note again that $|\alpha|^2$ corresponds to the photon flux and not the actual photon number.

In Appendix \ref{app:3}, we additionally provide an analytical solution for the continuous mean field approach 
using the Laplace transformation.
\begin{figure*}[t!]
\includegraphics[width= 0.49\textwidth]{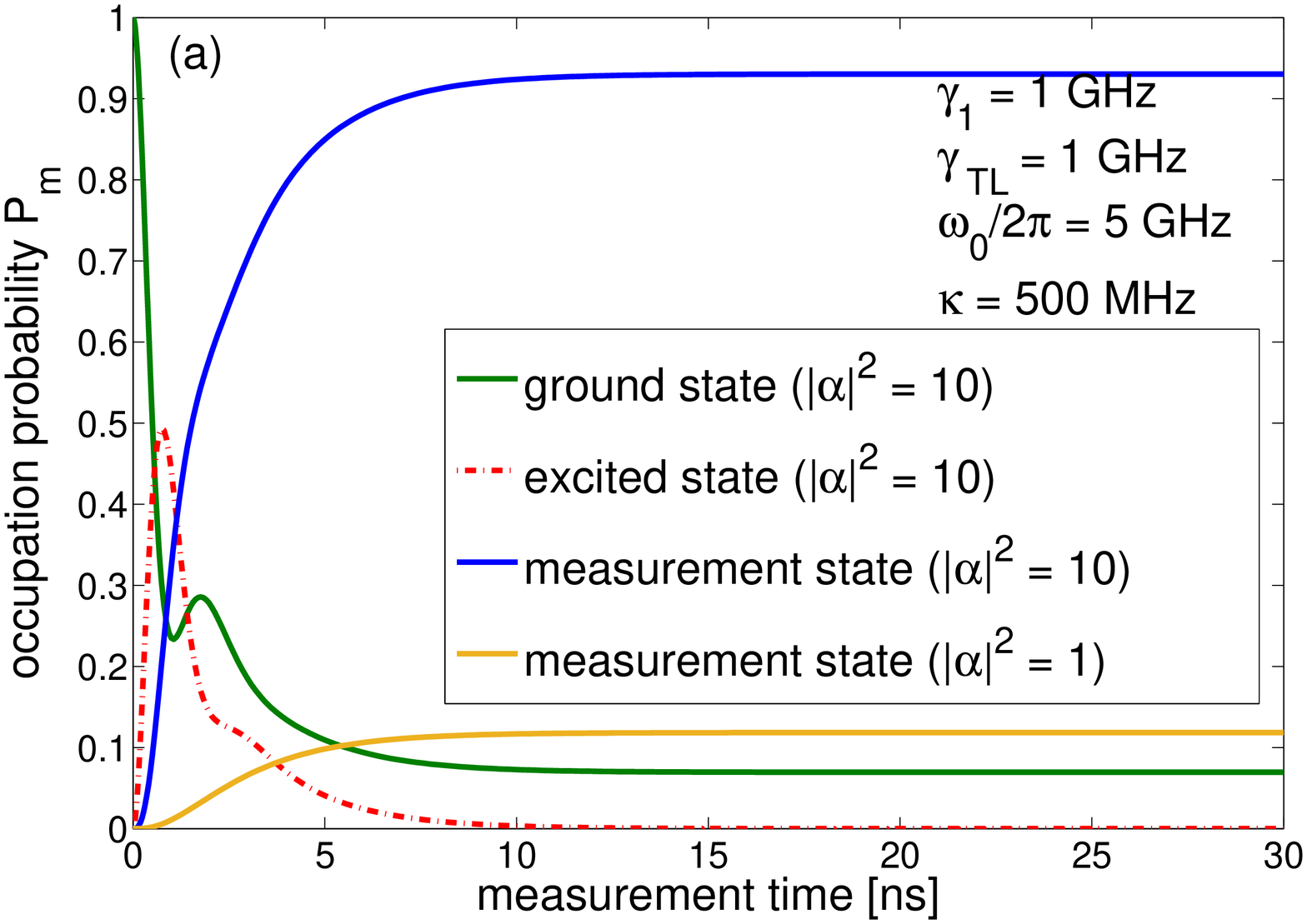}
\includegraphics[width= 0.49\textwidth]{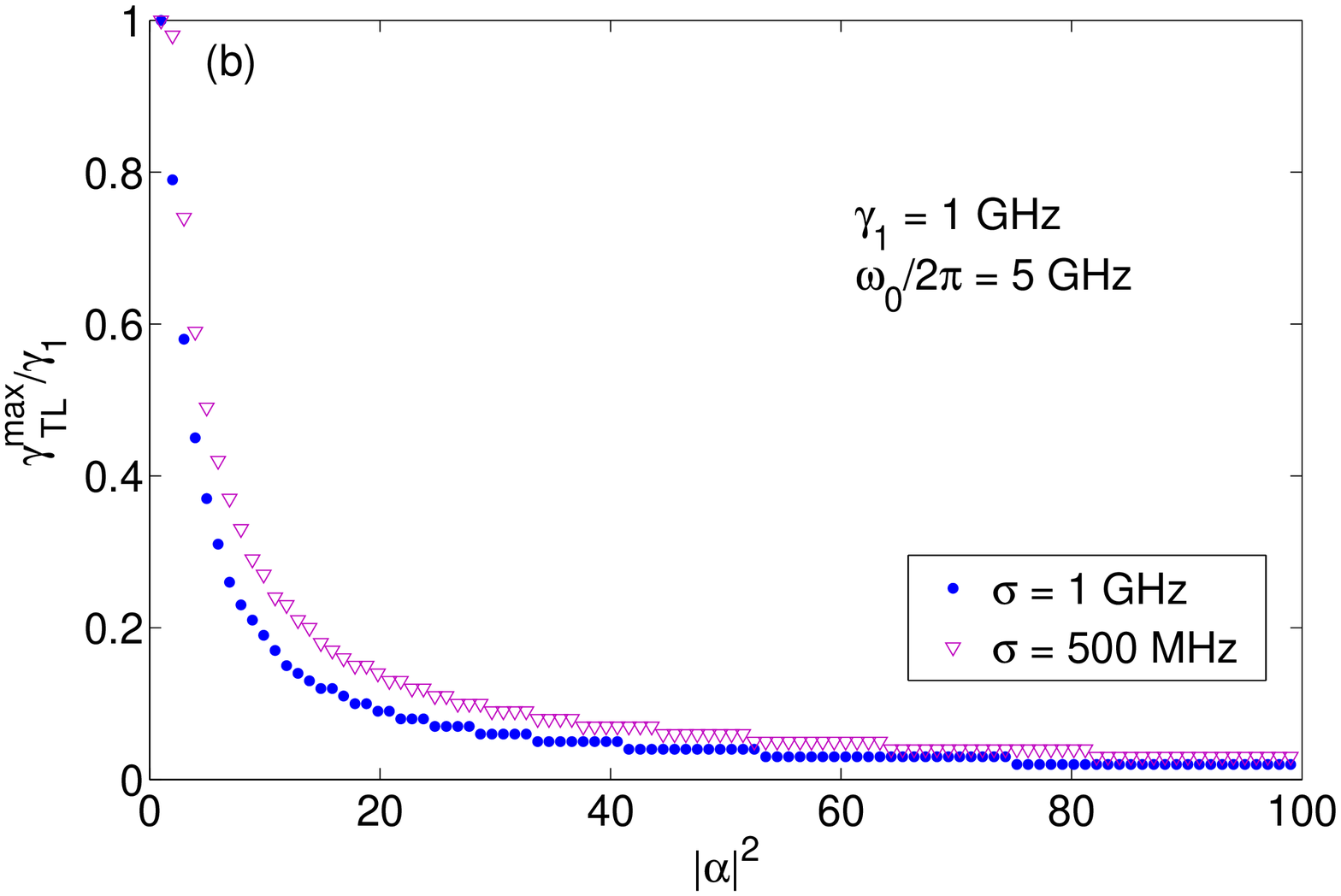}
\includegraphics[width=0.49\textwidth]{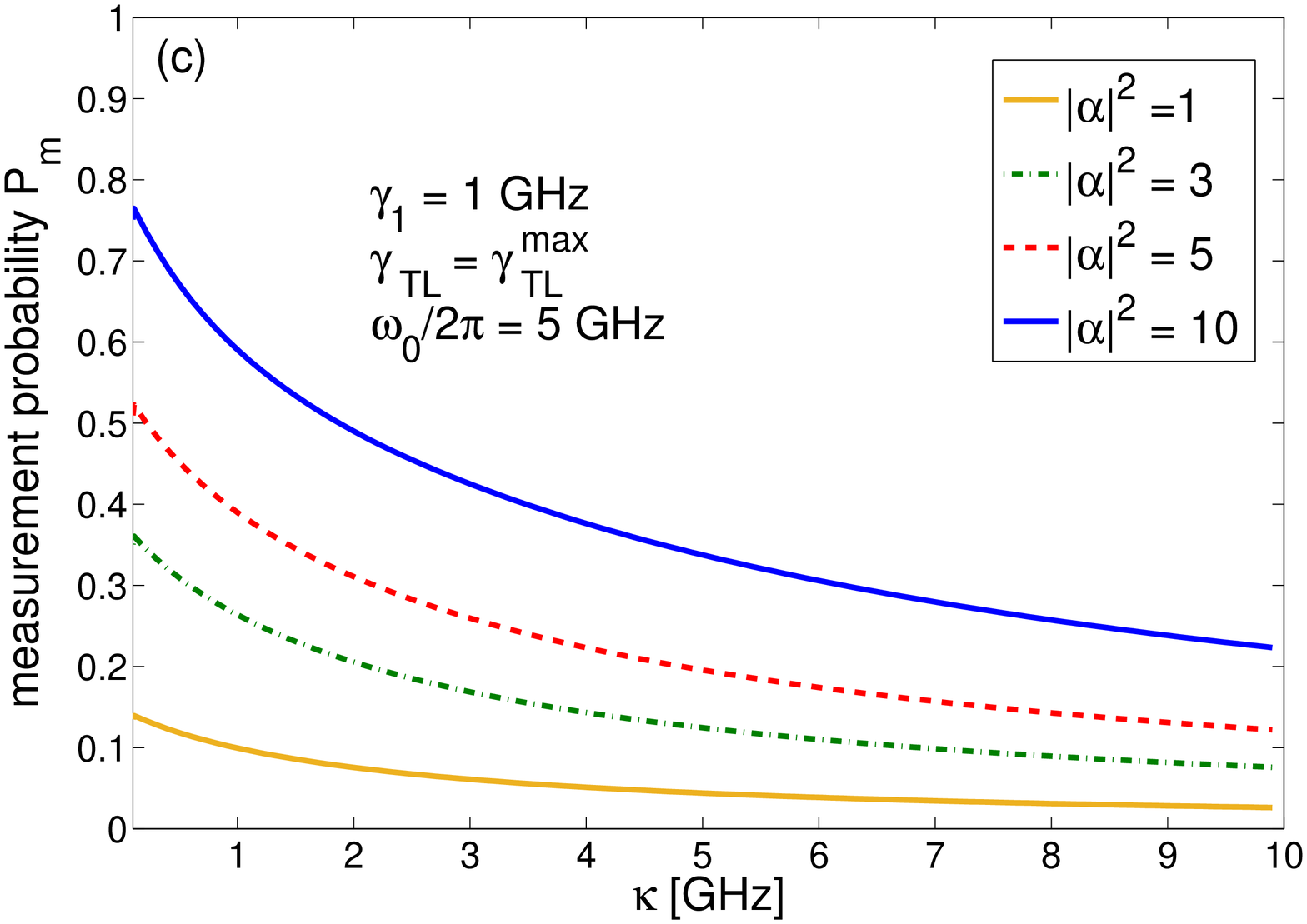}
\includegraphics[width=0.49\textwidth]{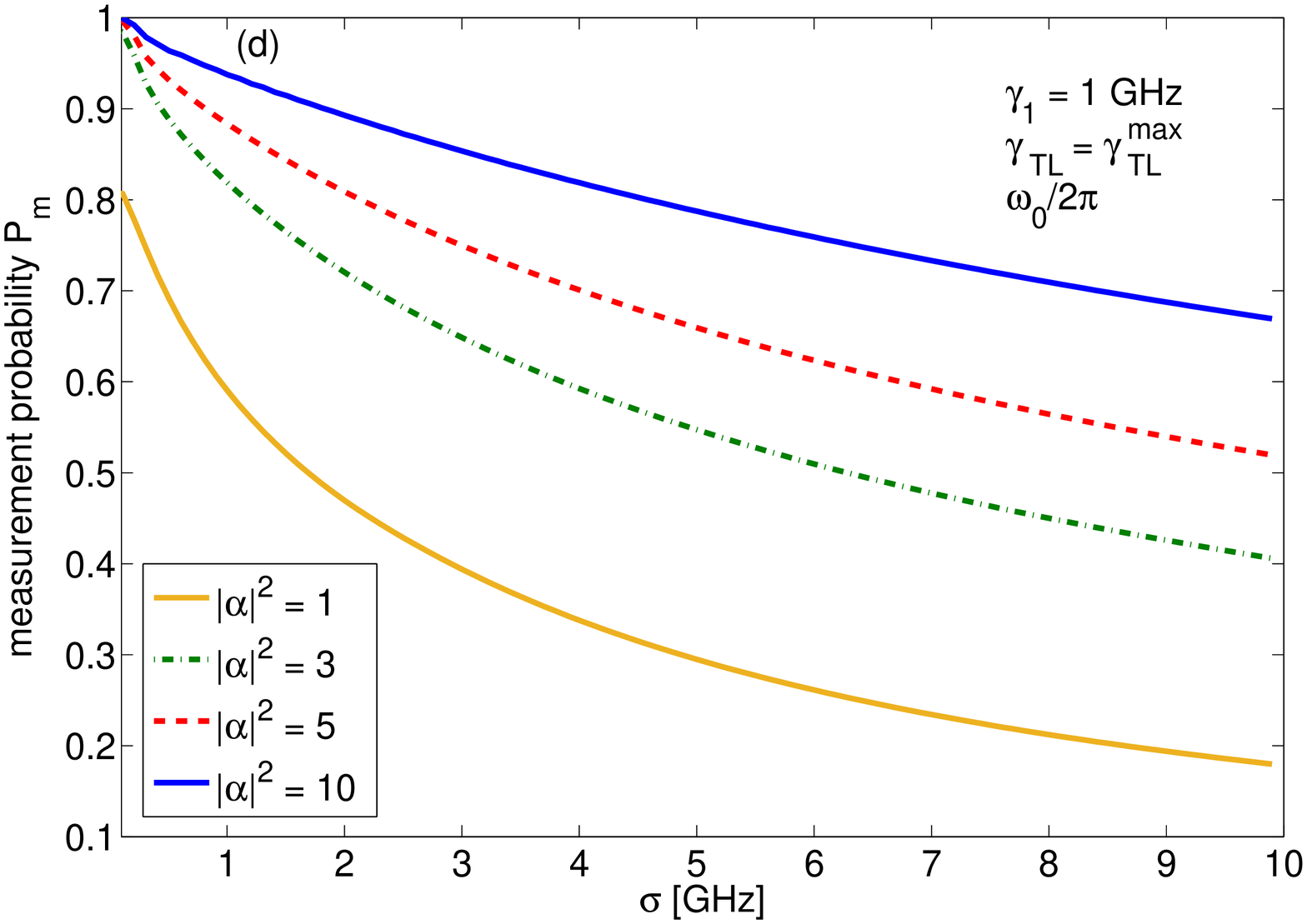}
\includegraphics[width=0.49\textwidth]{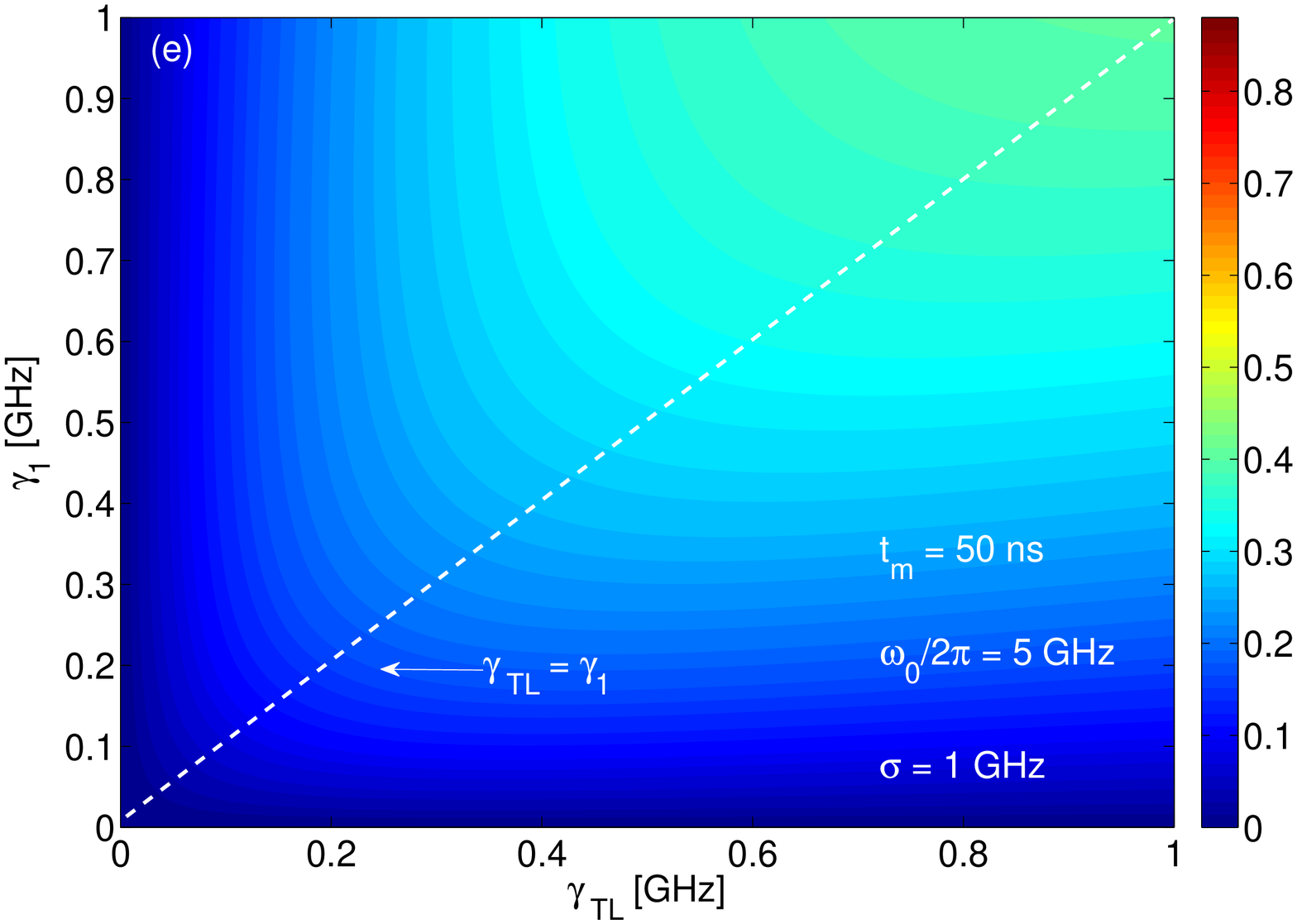}
\includegraphics[width=0.49\textwidth]{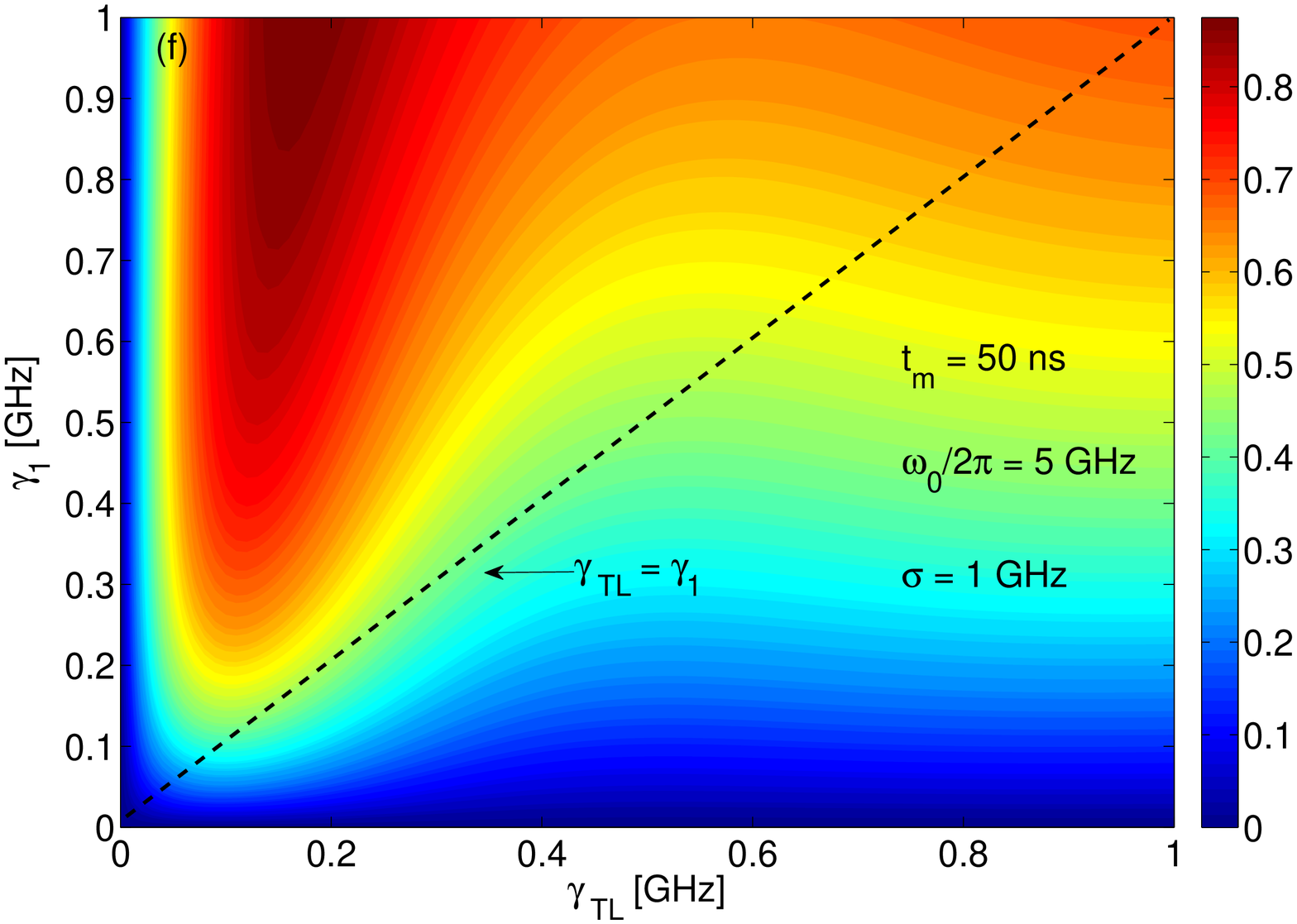}
\caption{\label{fig:gauss} Results for pulse shaped inputs. (a) Time evolution of the state occupation probabilities for an exponentially damped pulse with mean photon number $|\alpha|^2$. Additionally we show the measurement probability for  $|\alpha|^2 = 1$. (b) Dependence of the optimal choice of rates $\gamma_{\rm TL}^{\rm max}/\gamma_1$ on $|\alpha|^2$ for a Gaussian pulse. (c),(d) Dependence of the optimal measurement probability depending on $\kappa$ and $\sigma$ for exponentially damped and Gaussian pulses, respectively. Note that the x-axis does not start at $0$ since the pulse is not well defined for $\kappa = 0$ and $\sigma = 0$, respectively. (e), (f) Shift of the optimal measurement region for different values of $|\alpha|^2$ in the Gaussian case. (e) shows the behavior for $250$ photons arriving during $t_m$ and (f) for $2500$ photons arriving during $t_m.$}
\end{figure*}

\subsection{Pulsed Input}

For applications to qubit measurement \cite{govia2014high} we wish to perform threshold detection on a \textit{coherent} input pulse of $n$ photons. Therefore, we want to extend the above solutions to the more general case of an arbitrary input waveform. In this case the form factor $f(\omega)$ is no longer proportional to a simple $\delta$-function, it describes the shape of the pulse in the frequency space. We assume the form factor in the time domain $f(t)$, which is given by the Fourier transformation of $f(\omega)$, to be real. 

Note that especially in quantum optical treatments it is typical to include additional noise operators into the ladder operators, since they treat noise channels as additional input/output fields. However, we include all noise channels directly through Lindblad operators and therefore have no need to include additional noise channels in the expression for $\hat a_{\rm out}$ and $\hat a_{\rm in}$. We incorporate this form factor into the system of equations and follow the same procedure as in the previous section.

By using the Fourier relation $\int_{-\infty}^{\infty}{\rm d}\omega f(\pm\omega) \exp\left(\mp i \omega t\right) = f(t)$ we can bring the resulting system of equations to the following form:
\begin{subequations}
\begin{align}
\label{Pulse1}
\left<\dot{\hat \sigma}^{-}\right> &= - \frac{\tilde \gamma}{2} \left<\hat \sigma^{-}\right>  - i \frac{\omega_R(t)}{2}   \left(\left<\mathcal{\hat P}_0\right> - \left<\mathcal{\hat P}_1\right>\right) \\ 
\label{Pulse2}
\left<\dot{\hat \sigma}^{+}\right> &= - \frac{\tilde \gamma}{2} \left<\hat \sigma^{+}\right>+   i \frac{\omega_R(t)}{2}   \left(\left<\mathcal{\hat P}_0\right> - \left<\mathcal{\hat P}_1\right>\right)    \\
\label{Pulse3}
\left<\dot{\hat{\mathcal P}}_0\right> &= \gamma_{\rm TL} \left<\mathcal{\hat P}_1\right> - i \frac{\omega_R(t)}{2}  \left(\left<\hat \sigma^{-}\right>  - \left<\hat \sigma^{+}\right> \right)   \\
\label{Pulse4}
\left<\dot{\hat{\mathcal P}}_1\right> &= -\tilde \gamma \left<\mathcal{\hat P}_1\right> + i \frac{\omega_R(t)}{2}  \left(\left<\hat \sigma^{-}\right>  - \left<\hat \sigma^{+}\right> \right)    \\
\label{Pulse5}
\left<\dot{\hat{\mathcal P}}_m\right> &= \gamma_1 \left<\mathcal{\hat P}_1\right>,
\end{align} 
\end{subequations}
where $\omega_R(t) \equiv f(t)\sqrt{2|\alpha|^2\gamma_{\rm TL}/\pi}$ depends on the pulse shape in the time domain. This system of equations is similar to that in the previous section, apart from an additional factor $f(t)$ that specifies the pulse shape. Using these equations, we can solve for the time evolution of the state occupations for an arbitrary pulse shape.

Here, we study two different shapes, an exponential damped pulse and a Gaussian pulse.
The first pulse shape is especially relevant for qubit measurement, since it describes the shape of a pulse created from a spontaneous emission source \cite{meschede1985one,brune1987realization,ginzel1993quantum,wenner2014catching}. This pulse is described by the form factor
\begin{align}
 f(t) &= \sqrt{\kappa}\exp\left(-\frac{\kappa}{2}t\right)
\end{align}
with signal frequency $\omega_s$ of the pulse and duration $\tau_e = 2\pi/\kappa$.  Again we assume the signal frequency to be equal to the JPM transition frequency, $\omega_s = \omega_0$.

Next, we study the most natural choice for a few-photon wave packet, namely the Gaussian pulse
\begin{align}
\begin{split}
 f(\omega) &= \frac{1}{\left(2\pi \sigma^2\right)^{\frac{1}{4}}} \exp\left(-\frac{(\omega-\omega_s)^2}{4\sigma^2}\right) \\  f(t) &= \left(8\pi\sigma^2\right)^{\frac{1}{4}} \exp\left(-\sigma^2 (t-t_0)^2\right),
 \end{split}
\end{align} 
with duration $\tau_G = 2\pi/\sigma$. We assume the signal frequency $\omega_s$ to coincide with the transition frequency of the JPM ($\omega_s = \omega_0$). Note that we choose $t_0$ different from zero to include all of the Gaussian features (i.e. choose $t_0$ such that both minima of the pulse are included). The results are similar to the results for the exponentially damped pulse, except that $\sigma$ plays the role of $\kappa$ in this case (see Fig. \ref{fig:gauss}). Note that all the pulses are normalized to one, which means $\int_{0}^{\infty} {\rm d}t |f(t)|^2=1$.

For small amplitudes $|\alpha|^2$, we observe the matching condition \eqref{matching_simple} we found in Sec. \ref{sec:4A}. Increasing $|\alpha|^2$ shifts the maximum regime to higher values of $\gamma_1$ and smaller values of $\gamma_{\rm TL}$, for the same reason as in the continuous drive case. The behavior of $\gamma_{\rm TL}^{\rm max}/\gamma_1$ for a Gaussian pulse is shown in Fig. \ref{fig:gauss}(b) for two different values of $\sigma$. The agreement between the matching condition in the continuous case and the pulse case can be explained by the fact that a continuous drive is a special case of e.g a Gaussian pulse when $\sigma \longrightarrow 0$. Therefore it makes sense that we found the same optimization conditions at least for small enough $\sigma$. Anyways Fig. \ref{fig:gauss} indicates that the agreement can also be found for higher values of $\sigma$. We see that the ratio starts at one and then immediately drops to smaller values before asymptotically tending to zero in the classical regime. The movement of the optimal measurement region is also shown in Fig. \ref{fig:gauss}(e-f). In contrast to the continuous drive case, the measurement probability for pulsed input does not saturate at one, since a finite number of photons hits the detector. The actual value of $P_m$ in the steady state depends heavily on $|\alpha|^2$ (see Fig. \ref{fig:gauss}(a)) 

On the other hand, the maximum of the measurement probability for fixed values of $|\alpha|^2$ depends on the parameters $\kappa$ and $\sigma$ for the exponentially damped and Gaussian pulse, respectively (see Fig. \ref{fig:gauss}(c),(d)). In both cases we see that the shorter the pulse, the smaller the measurement probability since for longer pulses it is more likely that a photon excites the JPM.  

For the exponentially damped pulse, it is also possible to obtain analytical results using the Laplace transformation. We find the following expression for the measurement probability in the stationary state

\begin{align}
\begin{split}
\lim\limits_{t \rightarrow \infty} \left<\mathcal{\hat P}_m(t)\right> &= \frac{\tilde \omega_R^2}{4\kappa\left(\kappa+\frac{\tilde \gamma}{2}\right)\left(1 + \frac{\gamma_{\rm TL}}{\gamma_1}\right)} \\&\hspace{-0.5cm}- \sum_{l=0}^{\infty} \frac{\omega_R^2}{2} \frac{1+4\frac{\kappa}{\gamma_1}}{\left(\kappa+\frac{\tilde \gamma}{2}\right)\left(1+\frac{\gamma_{\rm TL}}{\gamma_1}\right)} \frac{\left<\mathcal{\hat P}_m(0)^{(l)}\right>}{(2\kappa)^{-(l+1)}}.
\end{split}
\end{align}
with $\tilde\omega_R = \sqrt{2|\alpha|^2\kappa\gamma_{\rm TL}/\pi}$. For given initial conditions of the system and starting with the set of equations \eqref{Pulse1}-\eqref{Pulse5}, one can calculate $\left<\mathcal{\hat P}_m(0)^{(l)}\right>$ to arbitrary order (for more details see Appendix \ref{app:4}). In Fig \ref{fig:analytic_comparison} we see the deviation between the the analytic solution up to fifth order and the numerical solution of \eqref{Pulse1}-\eqref{Pulse5}. We see that increasing $\alpha$ and small $\kappa$ the devitation is quite high, since the small parameter is $\alpha/\kappa$, but for high kappa the agreement is very good. 
\begin{figure}
\includegraphics[width=0.5\textwidth]{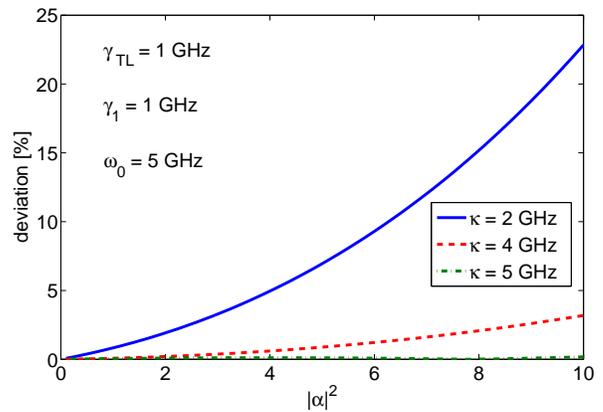}
\caption{\label{fig:analytic_comparison} Deviation of the steady state measurement probability between the analytical solution up to fifth order and the numerical solution for the exponentially damped pulse with different values of $\kappa$, as a function of $|\alpha|^2$. For small $\kappa$ and large $|\alpha|^2$ the deviation is quite high, but for increasing $\kappa$ the approximation fits the numerical results well. For $\kappa = 5$ GHz the deviation is almost zero.}
\end{figure}

\section{Rate equations}
\label{sec:4}

In this section, we approximate equations \eqref{System1}-\eqref{System5} and find optimal conditions to maximize measurement efficiency in the stationary state. We assume $\gamma_{\rm res} \neq 0$ to derive rate equations for the occupation probabilities which can be solved analytically. The measurement probability is then given by the occupation probability of the measurement state. In case of the JPM it is difficult to reset the counter since it tunnels into a continuum of states. Ideas exist to reset the JPM using relaxation oscillations, but for the time being the JPM is restricted to a single measurement. For this reason we assumed the reset rate to be zero in Sec. \ref{sec:3}. However, our techniques are general, and can be applied to any counter, e.g. a counter based on a driven $\Lambda$ system \cite{inomata2016single} can be reseted using a control pulse that drives the system back to its initial state. For such a system the reset times are around $400$ ns.  

The main result of this section will be an analytical derivation of the matching condition for small input fields that was found in the last section. We also derive a generalized matching condition where we include dark counts and relaxation.  All results in this section are for a continuous drive, as we cannot treat pulses with this approach. Additionally, we extend the results to the case where dark counts $(\gamma_0\neq 0)$ and relaxation processes $(\gamma_{\rm rel}\neq 0)$ are present.

\begin{figure*}[t!]
\includegraphics[width=0.49\textwidth]{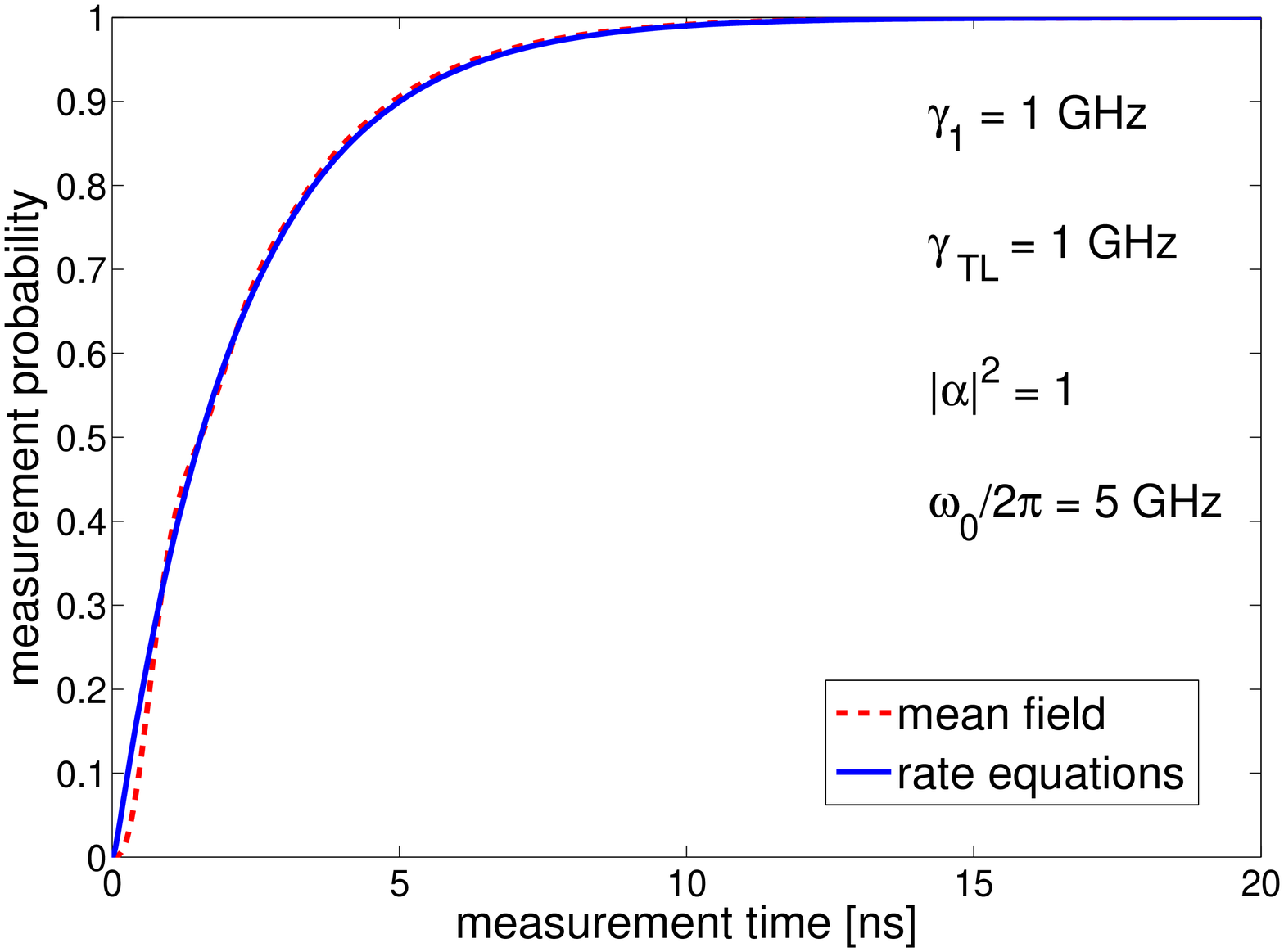}
\includegraphics[width=0.49\textwidth]{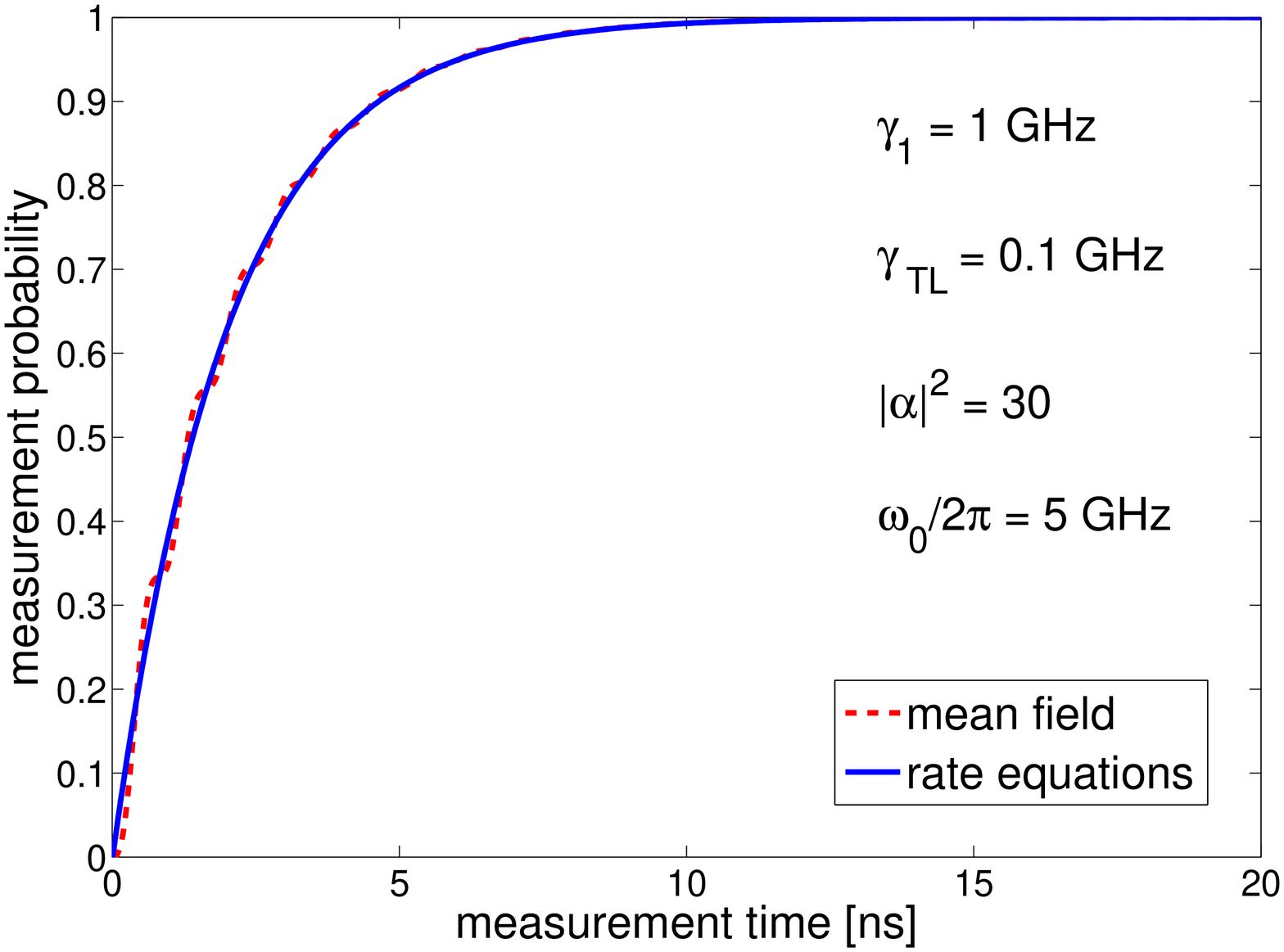}
\caption{\label{fig:comparison} Comparison of measurement probability given by the numerical solution of equation system \eqref{continuous_1}-\eqref{continuous_5} (red) and the analytical solution of the rate equations \eqref{Gl:A2} found in App. \ref{app:2} (blue), in the quantum (left) and classical regimes (right). The two approaches give similar results apart from the absence of Rabi oscillations in the rate equation approach, where the JPM is treated classically.}
\end{figure*}

The approach used in this section to derive the rate equations obscure the quantum mechanical nature of the system, and do not capture effects such as the Rabi oscillations in the measurement probability. The missing Rabi oscillations can probably be explained by the fact, that we ignore correlations between the field and the JPM due to approximation \eqref{approx_rate} (wit this approximation we automatically split expectation values like $\left<\mathcal{\hat O}_{0} \hat a_{\rm in}^{\dag}\right> \approx \left<\mathcal{\hat O}_{0} \right>\left<\hat a_{\rm in}^{\dag}\right>$). However, the results of this section still coincide well with the results of the mean field approach, especially the average measurement probability (see Fig. \ref{fig:comparison}).

In the limit of fast decay of $\hat \sigma_z$, we can assume that the JPM dynamics are entirely incoherent (i.e. the expectation values of $\hat \sigma_x$ and $\hat \sigma_y$ decay quickly), hence we substitute for the operator $\hat \sigma_z$ its expectation value
\begin{align}
\hat \sigma_z(t) \longmapsto \left<\sigma_z(t)\right> = P_0(t)-P_1(t),
\label{approx_rate}
\end{align}
where $P_0$ and $P_1$ denote the probability to be in the ground and excited state, respectively. Given the many rates contributing to the decay of $\hat \sigma_z$, this condition is met under a wide range of parameters, consistent with the effectiveness of this approximation that we shall demonstrate later on (see Fig. \ref{fig:comparison}). Especially the rate $\gamma_1$ should be large in experiment, since it determines how fast the tunneling from the metastable state into the measurement state happens.

We want to study a continuous resonant drive $\omega_s = \omega_0$, such that 
\begin{align}
f(\omega)\hat a_{t_0}(\omega) =  \sqrt{\omega_0}\delta(\omega-\omega_0) \hat a_{t_0}(\omega),
\label{incoming_radiation}
\end{align}
similar to Sec. \ref{sec:4A}. In this case, the Fourier transformation of \eqref{System1} can easily be done:
\begin{align}
\begin{split}
-i \omega_0 \hat  \sigma^{-}(\omega_0) = &-\left(i\omega_0+\frac{\tilde \gamma}{2}\right)\hat \sigma^{-}(\omega_0) \\ &+ \sqrt{\gamma_{\rm TL}} \hat a_{\rm in}(\omega_0)(P_0(\omega_0)-P_1(\omega_0)).
\end{split}
\label{help1}
\end{align}
Note that all the appearing operators actually act on the transmission line and the JPM. While e.g. $\hat \sigma^{-}(t=0)=\hat \sigma^{-}\otimes\mathbbm{1}$ acts as the identity on the transmission line, this is no longer true at later times, highlighting the build-up of entanglement. One nicely sees this in Eq. \eqref{help1}, the second part of the right hand side leads to a contribution to $\hat \sigma^{-}$ that acts on the transmission line, i.e. if the JPM is in the ground state, then $\hat \sigma^{-}$ becomes more transmission line like as time evolves, hence the plus sign, while if the JPM is in the excited state the qubit operator becomes less transmission line like, hence the minus sign. Here transmission line like corresponds to the field operator part $\hat \sigma^{-}$ gets due to the time evolution under Eq. \eqref{help1}. 
With \eqref{help1}, relation \eqref{inout_Relation} leads to 
\begin{align}
\hat a_{\rm out}(\omega_0) = R(\omega_0) \hat a_{\rm in}(\omega_0),
\label{Reflection_1}
\end{align}
with the reflection coefficient
\begin{align}
R(\omega_0)   = -1 + \frac{2\gamma_{\rm TL}}{\tilde \gamma} \left[P_0(\omega_0)-P_1(\omega_0)\right].
\label{Reflection_omega}
\end{align}
Inverse Fourier transform of Equation \eqref{Reflection_1} yields the time-domain relation 
\begin{align}
\hat a_{\rm out}(t) &= \mathcal{F}^{-1}\left[R(\omega_0)\right] \ast \mathcal{F}^{-1}\left[\hat a_{\rm in}(\omega_0)\right] \\
&= R(t) \hat a_{\rm in}(t),
\label{help2}
\end{align}
where we have only to substitute $P_{0/1} (\omega_0)$ with $P_{0/1}(t)$ in Equation \eqref{Reflection_omega} for $R$, because $\hat a_{\rm in}\propto \delta(\omega-\omega_0)$, which makes the resulting convolution easy to solve. Note that this is only possible for a continuous drive. The absolute value of the reflection coefficient in our system can be greater than one if $P_0(t)<P_1(t)$, because in this case the incoming signal can be amplified by spontaneous or stimulated emission. All the equations \eqref{help1}-\eqref{help2} are also valid for the non-resonant case, provided one substitutes $\omega_0$ in \eqref{incoming_radiation} with a frequency that is not equal with the JPM transition frequency $\omega_s \neq \omega_0$.

To obtain the rate equations for the system, we replace $\mathcal{\hat P}_0$, $\mathcal{\hat P}_1$, and $\mathcal{\hat P}_m$ with the corresponding occupation probabilities $P_0$, $P_1$, and $P_m$, which leads to
\begin{subequations}
\begin{align}
\dot{P}_0 &= -\gamma_0 P_0 + (\gamma_{\rm TL}+\gamma_{\rm rel})P_1 \\ &\hspace{0.5cm}-\sqrt{\gamma_{\rm TL}}\left(\left<\hat{a}_{\rm in}^{\dagger} \hat \sigma^{-} \right> + \left<\hat \sigma^{+}\hat{a}_{\rm in}\right>\right) + \gamma_{\rm res} P_m \nonumber\\
\dot{P}_1 &= -(\gamma_{\rm TL}+\gamma_1+\gamma_{\rm rel})P_1 +\sqrt{\gamma_{\rm TL}}\left(\left<\hat{a}_{\rm in}^{\dagger}\hat \sigma^{-} \right>  + \left<\hat \sigma^{+}\hat{a}_{\rm in} \right>\right)\\
\dot{P}_m &= \gamma_0 P_0 + \gamma_{1} P_1 -\gamma_{\rm res} P_m.
\end{align}
\end{subequations}
Using relation \eqref{inout_Relation} and the expression for $R$, we end up with a system of coupled rate equations where we have eliminated $\hat \sigma^{-}$ and $\hat \sigma^{+}$
\begin{subequations}
\begin{align}
\dot{P}_0 &= -(\beta N_{\rm in}+\gamma_0) P_0 +(\beta N_{\rm in}+\gamma_{\rm TL}+\gamma_{\rm rel})P_1 + \gamma_{\rm res} P_m \label{rate_1}\\
\dot{P}_1 &= \beta N_{\rm in} P_0 -(\beta N_{\rm in}+\gamma_{\rm TL}+\gamma_1+\gamma_{\rm rel}) P_1, \label{rate_2}\\
\dot{P}_m &= \gamma_0 P_0 + \gamma_1 P_1 - \gamma_{\rm res} P_m \label{rate_3}.
\end{align}
\end{subequations}
with $\beta = \frac{2}{\pi}\frac{\gamma_{\rm TL}}{\tilde \gamma}$
and the incoming photon flux $N_{\rm in} = \left<\hat a^{\dag}\hat a\right>$ (for more details see Appendix \ref{app:photon_flux}).
Note that $\beta < 1$, such that the excitation rate of ground to excited state is smaller than the rate of incoming photons. 

The overall measurement efficiency is given in the stationary state; therefore, we set $\dot P_0 = \dot P_1 = \dot P_m = 0$. Doing so and using the constraint $P_0 + P_1 + P_m = 1$, we end up with an expression for stationary $P_0$, and $P_1$
\begin{align}
\begin{split}
P_0&= \frac{1}{1+\frac{\gamma_0}{\gamma_{\rm res}}} \\&\hspace{-0.3cm}- \frac{\beta \tilde \gamma N_{\rm in}}{\tilde \gamma \left(\beta N_{\rm in} \left[\frac{\gamma_{\rm res}+\gamma_1}{\gamma_{\rm res}+\gamma_0}\right]+\gamma_{\rm TL}+\gamma_1+\gamma_{\rm rel}\right)\left(1+\frac{\gamma_0}{\gamma_{\rm res}}\right)^2}
\end{split}
\label{rate_p0}
\\
P_1&= \frac{\beta \tilde\gamma N_{\rm in}}{\tilde \gamma \left(\beta N_{\rm in} \left[\frac{\gamma_{\rm res}+\gamma_1}{\gamma_{\rm res}+\gamma_0}\right]+\gamma_{\rm TL}+\gamma_1+\gamma_{\rm rel}\right)\left(1+\frac{\gamma_0}{\gamma_{\rm res}}\right)}. 
\label{rate_p1}
\end{align}
To get from equations \eqref{rate_1}-\eqref{rate_3} to the expressions \eqref{rate_p0}, \eqref{rate_p1} we had to assume that $\gamma_{\rm res}>0$, such that the expressions for $P_0$ and $P_1$ are only valid for the case $\gamma_{\rm res}\neq 0$. The exact solution for the case $\gamma_{\rm res}=0$ is given in Appendix \ref{app:2}.

\begin{figure}
\includegraphics[width= 0.5\textwidth]{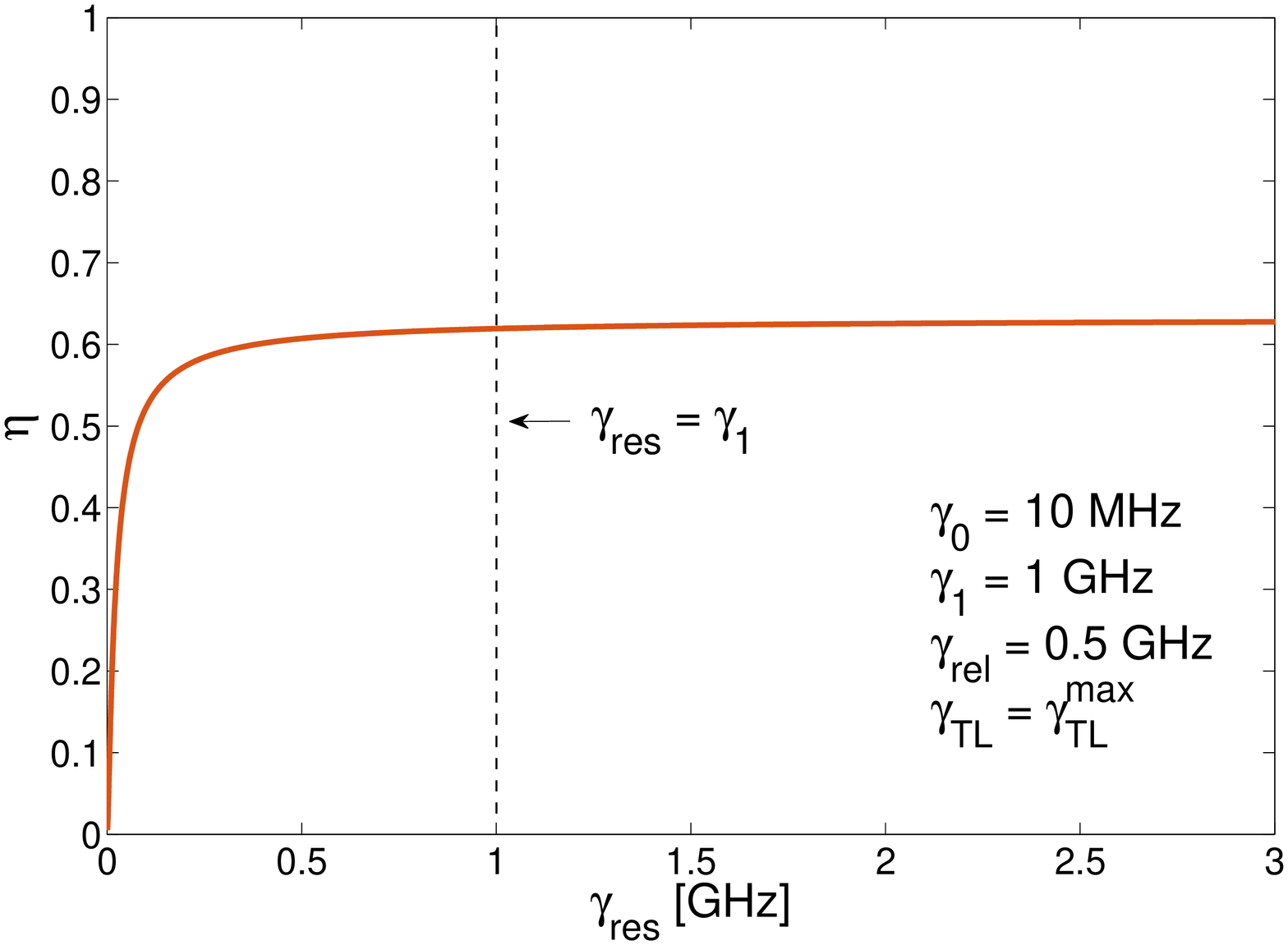}
\caption{\label{fig:efficiency2} Efficiency $\eta$ as a function of the reset rate $\gamma_{\rm res}$. For small values of $\gamma_{\rm res}$, increasing the reset rate leads to a strong enhancement of the efficiency up to a point where the reset is roughly as fast as the decay into the measurement state ($ \gamma_{\rm res} \approx \gamma_1$). From then on the efficiency stays constant with increasing $\gamma_{\rm res}$, since the reset is faster than the average measurement time.}
\end{figure}

The dark count correction is given by the counting rate in absence of incoming photons; therefore, $\Gamma_{\rm dark} = \gamma_0P_0(N_{\rm in}=0)$. If we use the fact that the dead time of the counter can be expressed in terms of the reset rate as $\tau_{\rm dead} = 1/\gamma_{\rm res}$, we obtain the well known expression for the dark count correction for quantum optical counters \cite{hadfield2009single}
\begin{align}
\Gamma_{\rm dark} = \gamma_0 P_0(N_{\rm in}=0) = \frac{\gamma_0}{1+\gamma_0\tau_{\rm dead}}.
\label{dark}
\end{align}
The overall counting rate on the other hand is given by
\begin{align}
\Gamma_{\rm count}  = \gamma_1 P_1(N_{\rm in})+\gamma_0 P_0(N_{\rm in}).
\label{bright}
\end{align}
With \eqref{dark} and \eqref{bright}, the bright count rate, which describes the rate at which incoming photons are detected, can be written as
\begin{align}
\Gamma_{\rm bright} = \Gamma_{\rm count} - \Gamma_{\rm dark}.
\end{align}
The fidelity of a photon counter can in general be characterized by its efficiency, which is defined as the rate of detected photons $\Gamma_{\rm bright}$ over the rate of incident photons $\Gamma_{\rm incident} = N_{\rm in}$ \cite{hadfield2009single}. For the JPM, the efficiency is given by
\begin{align}
\eta  &= \frac{\Gamma_{\rm bright}}{\Gamma_{\rm incident}} \label{eff1}  \\
     &= \frac{1}{N_{\rm in}} \left[\gamma_1 P_1(N_{\rm in}) + \gamma_0 P_0(N_{\rm in}) - \gamma_0P_0(N_{\rm in}=0)\right].\nonumber
\end{align}
If we put the expressions for $P_0$ and $P_1$ into \eqref{eff1}, we obtain an overall expression for the detection efficiency:
\begin{align}
\eta = \frac{4 \gamma_{\rm TL}\gamma_{\rm res} \left[\gamma_1\left(\gamma_0+\gamma_{\rm res}\right)+ \gamma_0 \left(\gamma_1+\gamma_{\rm res}\right)\right]}{(\gamma_{\rm TL}+\gamma_1+\gamma_{\rm rel})(\gamma_{\rm TL}+\gamma_1+\gamma_0+\gamma_{\rm rel})\left(\gamma_0+\gamma_{\rm res}\right)^2},
\label{efficiency_general}
\end{align}
 where we have assumed the low excitation limit ($N_{\rm in}/\omega_0\ll1$), such that the terms proportional to $N_{\rm in}$ in the denominators of \eqref{rate_p0} and \eqref{rate_p1} can be ignored.
\begin{figure}
\includegraphics[width= 0.5\textwidth]{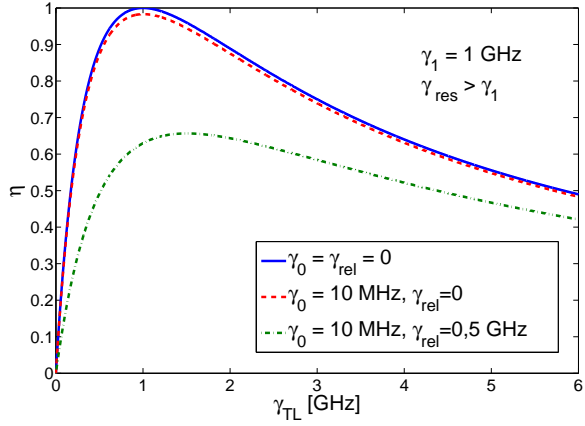}
\caption{\label{fig:rate_approach} Efficiency  $\eta$ as a function of the coupling rate $\gamma_{\rm TL}$. The efficiency has a distinct maximum value given by equation \eqref{matching_general} that depends on $\gamma_0$, $\gamma_1$, $\gamma_{\rm rel}$. For $\gamma_0 = \gamma_{\rm rel} = 0$ (blue), the general matching condition simplifies to \eqref{matching_simple} also found in the last section and the efficiency reaches 1. An additional dark count rate $\gamma_0$ (red) leads to a small shift and reduction of the maximum value; both are barely visible for typical values of $\gamma_0$. On the other hand, the inclusion of relaxation $\gamma_{\rm rel}$ (green) reduces the maximum value significantly and furthermore leads to a visible shift of the maximum to higher values of $\gamma_{\rm TL}$.} 
\end{figure} 
The efficiency possesses a distinct maximum (see Fig. \ref{fig:rate_approach}) that is reached when the following relation between rates is satisfied
\begin{align}
\gamma_{\rm TL}^{\rm max} = \sqrt{(\gamma_1+\gamma_{\rm rel})(\gamma_1+\gamma_{\rm rel}+\gamma_0)}.
\label{matching_general}
\end{align}
We refer to this expression as the general matching condition, since compared to \eqref{matching_simple} it additionally includes dark counts and relaxation. Note that the matching condition itself does not depend on $\gamma_{\rm res}$, but if $\gamma_{\rm res} < \gamma_1$ it limits the maximal efficiency (see Fig. \ref{fig:efficiency2}). If the rates are chosen such that \eqref{matching_general} is satisfied, we say the JPM and the transmission line are matched, to make a connection to impedance matching in microwave circuits \cite{Pozar}. When the JPM is matched to the transmission line and under the condition $\gamma_{\rm res}>\gamma_1$, we find an efficiency
\begin{align}
\eta_{\rm max} &= \frac{4(\gamma_0+\gamma_1)}{\gamma_0+2\left(\gamma_1+\gamma_{\rm rel} + \sqrt{(\gamma_1+\gamma_{\rm rel})(\gamma_0+\gamma_1+\gamma_{\rm rel})}\right)},
\label{rate_efficiency}
\end{align}
To get expression \eqref{rate_efficiency} out of \eqref{efficiency_general} we assumed a high reset rate $\gamma_{\rm res}\gg \gamma_1$, hence $\gamma_1/\gamma_{\rm res}\approx 0$. However Fig. \ref{fig:efficiency2} indicates that \eqref{rate_efficiency} is valid as soon as $\gamma_{res}$ exceeds $\gamma_1$. 
If there are no dark counts and no relaxation, the efficiency is given by
\begin{align}
\eta = \frac{4\gamma_{\rm TL}\gamma_1}{(\gamma_{\rm TL}+\gamma_1)^2},
\label{eq:eff1}
\end{align}
and the general matching condition simplifies to the matching condition
\begin{align}
\gamma_{\rm TL} = \gamma_1,
\label{matching_simple}
\end{align}
that coincides with the result found in the last section.

This result coincides with the optimal matching condition found in Romero \textit{et al.} \cite{romero2009photodetection}; however, the efficiency was limited to $1/2$. The reason for this is that Romero \textit{et al.} assumed an infinite transmission line with a JPM in the middle. Therefore, an excitation in the JPM can spontaneously emit into the other side of the transmission line at a rate $\gamma_{\rm TL}$, allowing for transmission through the JPM. For maximum efficiency $\gamma_{\rm TL} = \gamma_1$, both photon detection and photon transmission through the JPM will occur with equal probability, reducing the efficiency to $1/2$. In this work, we assume a semi-infinite transmission line terminated by the JPM, such that the transmission process is not possible, which leads to a maximum efficiency of $1$.

In our case there are four main processes that limit detector efficiency: coupling losses (reflection), energy relaxation, dark counts, and dead time. Usually one distinguishes between two separate efficiencies: the efficiency due to coupling losses $\eta_{\rm loss}$ and the intrinsic quantum efficiency of the detector $\eta_{\rm det}$. Here, $\eta_{\rm loss}$ includes the effect of rate mismatch between the JPM and the transmission line, as described above. On the other hand, $\eta_{\rm det}$ includes the effects of dark counts, relaxation, and dead time. The overall efficiency can be written as the product of these two: $\eta = \eta_{\rm loss}\cdot \eta_{\rm det}$. 
Here $\eta_{\rm loss}$ can be extracted from \eqref{efficiency_general} by dividing it through \eqref{rate_efficiency}, since $\eta_{det} = \eta_{\rm max}$ (reflection losses are zero at matching point) and would in the general case (under the assumption $\gamma_{\rm res} \gg \gamma_1$) be given by
\begin{align}
\eta_{\rm loss} = \frac{\gamma_{\rm TL}\left(\gamma_0+2(\gamma_1+\gamma_0)+\sqrt{(\gamma_1+\gamma_{\rm rel})(\gamma_0+\gamma_1+\gamma_{\rm rel})}\right)}{(\gamma_{\rm TL}+\gamma_1+\gamma_{\rm rel})(\gamma_{\rm TL}+\gamma_0+\gamma_1+\gamma_{\rm rel})}
\end{align}
In the ideal case ($\gamma_0=\gamma_{\rm rel}= 0 $ and $\gamma_{\rm res} > \gamma_1$, such that $\eta_{\rm det}=1$), the efficiency is only limited by $\eta_{\rm loss}$. Condition \eqref{matching_simple} then determines the coupling rate for which coupling loss is zero, such that $\eta_{\rm loss} = 1$ and we reach unit efficiency (see Fig. \ref{fig:rate_approach}). This is exactly the point where all incoming photons reach the measurement state of the counter and all the incoming power is transferred into a measured signal. 

In the non-ideal case where we have dark counts and relaxation, even at the general matching point \eqref{matching_general} the efficiency is limited to a value smaller than one (since $\eta_{\rm det}<1$), such that the optimal power matching condition \eqref{matching_general} can only lead to an overall efficiency of $\eta_{\rm det}$ (see Fig. \ref{fig:rate_approach}).

In Fig. \ref{fig:efficiency2}, we see that the reset time also has a significant influence on $\eta_{\rm det}$. For $\gamma_{\rm res} < \gamma_1$, the efficiency increases rapidly with increasing $\gamma_{\rm res}$ up to the point where $\gamma_{\rm res} \approx \gamma_1$, after which the efficiency is approximately constant if we increase $\gamma_{\rm res}$. This can be explained by the fact that for a system with $\gamma_{\rm res} \approx \gamma_1$, the reset happens with the same rate as the measurement, such that increasing $\gamma_{\rm res}$ no longer has an influence on $\eta_{\rm det}$.
 
In many applications of detection of continuous-wave signals, it is helpful to express detector performance in terms of noise equivalent power (NEP), the effective noise power per unit bandwidth referred to the detector input. In the case of a photon counter with dark count rate $\gamma_0$ operated for an integration time $\tau$, Poisson uncertainty in the number of dark counts is given by $\sigma_N=\sqrt{\gamma_0\tau}$. Expressing this uncertainty as a photon flux at the input, we find (for the definition of the general NEP $\sigma_P$ see \cite{zmuidzinas2003thermal})
\begin{align}
\sigma_P = \frac{\hbar \omega_0}{\eta \tau} \sqrt{\gamma_0 \tau}.
\end{align}
If we choose an integration time of 0$.5$ s, corresponding to a detection bandwidth of $1$ Hz, we obtain the standard expression for the NEP of a photon counter \cite{zmuidzinas2003thermal, hadfield2009single}
\begin{align}
{\rm NEP} = \frac{\hbar\omega_0}{\eta}\sqrt{2\gamma_0};
\end{align}
if we put in the expression \eqref{efficiency_general} for JPM efficiency, we obtain the NEP for the JPM. For the JPM parameters $\gamma_{\rm rel} = 33$ kHz \cite{kelly2015state}, $\gamma_1 = 1$ GHz, $\gamma_0 = \gamma_1/100$ and $\omega_0/2\pi = 5$ GHz, we find an NEP of $2\times 10^{-20}$ W/$\sqrt{{\rm Hz}}$ at the matching point. This is to be compared against NEP of order $1 \times 10^{-17}$ W/$\sqrt{{\rm Hz}}$ achieved by transition edge sensors (TES) \cite{thornton2016atacama} and microwave kinetic inductance detectors (MKIDs) \cite{flanigan2016photon} at higher frequencies in the range from 40-300 GHz, relevant for cosmic microwave background (CMB) studies. 
It is possible that Josephson junctions based on higher-gap materials such as NbN could be used to realize JPMs with plasma frequencies in the tens of GHz range, suitable for low-noise detection of the CMB. 

In Appendix \ref{app:2}, we solve the time evolution of the rate equations of this section analytically to compare them to the results reached in Sec. \ref{sec:4A} for the continuous drive case. The comparison is shown in Fig. \ref{fig:comparison}. We see that the results of both approaches are very similar for both the classical and the quantum regimes.

Optimal conditions for photon detection using semiconductor quantum dots was also discussed in \cite{wong2015quantum}. In that case, the optimal condition is satisfied for the Cooperativity factor $\sim 1$.
\section{Conclusion}
\label{sec:5}

In conclusion, we have derived a general set of equations that describe a two-level photon counter strongly coupled to a transmission line. We have shown that one can reach high-efficiency photon detection of a traveling microwave state using appropriate matching of system parameters. The conditions vary for different input states; in general, for low input power the coupling rate between the counter and the transmission line should be equal to the measurement rate. At higher power, the matching condition shifts, such that the coupling rate should be smaller than the measurement rate. 

Because of the generality of the input-output formalism we used, the approach described here can be applied to arbitrary input pulses  and thus modified to fit the particular radiation source of any experiment. As a result, this work presents a guide to tune parameters to reach the optimal measurement efficiency for a range of experimental situations. Moreover, the presented method can be extended to any lossy two-level system coupled to a semi-infinite resonator.

\section*{Acknowledgements}

We thank Konstantin Nesterov, B. Taketani, Guilhem Ribeill, Ivan Pechenezhskiy and Ted Thorbeck for fruitful discussions.
Supported by the Army Research Office under contract W911NF-14-1-0080. LCGG and FKW also acknowledge support from the European Union through ScaleQIT and LCGG from NSERC through an NSERC PGS-D.

\begin{appendix}
\section{Hamiltonian of the system}
\label{app:1}

From the circuit diagram Fig. \ref{fig:JPM_TL} we can derive the Lagrangian of the system:
\begin{align}
\begin{split}
\mathcal{L} &= \mathcal{L}_{\rm TL} + E_J \cos(\varphi_J) + (I_b+\Delta I) \left(\frac{\Phi_0}{2\pi}\right) \varphi_J \\ &\hspace{1.1cm}+ \frac{1}{2} C_J \left(\frac{\Phi_0}{2\pi}\right)^2 \dot{\varphi}_J^2\\
&= \mathcal{L}_{\rm TL} + \mathcal{L}_{\rm JPM} + \Delta I \left(\frac{\Phi_0}{2\pi}\right) \varphi_J,
\label{eq:appendix1}
\end{split}
\end{align}
where $\mathcal{L}_{\rm TL}$ is the bare transmission line Lagrangian (sum of harmonic oscillators), $\varphi_J$  the phase of the JPM, $I_b$ the bias current, $E_J$ the Josephson energy, $C_J$ the junction capacitance, $\Phi_0$ the flux quantum, and $\Delta I$ the additional current coming from the transmission line. Here, $\mathcal{L}_{\rm JPM} \equiv E_J \cos(\varphi_J) + I_b \frac{\Phi_0}{2\pi} \varphi_J + \frac{1}{2} C_J \left(\frac{\Phi_0}{2\pi}\right)^2 \dot{\varphi}_J^2$ is the Lagrangian of the JPM. The last term of \eqref{eq:appendix1} leads to an interaction between the JPM and the transmission line. Using the Legendre transformation, we obtain the Hamiltonian of the system:
\begin{align}
\mathcal{H} = \mathcal{H}_{\rm TL}+\mathcal{H}_{\rm SYS} + \Delta I \frac{\Phi_0}{2\pi} \varphi_J,
\label{Hamilton_Appendix}
\end{align}
where $\mathcal{H}_{\rm TL}$ is the Hamiltonian describing the transmission line and $\mathcal{H}_{\rm JPM}$ is the Hamiltonian of the JPM.

We want to take a closer look at the interaction term. If we use the normal procedure of quantizing the transmission line and the JPM, we get the following expression for the current \cite{johansson2010dynamical} and phase operators \cite{geller2007quantum}:
\begin{align}
\Delta \hat I &= \sqrt{\frac{\hbar \omega_s}{4\pi Z_0}} \int_{0}^{\infty} {\rm d}\omega\left(\hat a^{\dagger}(\omega)+\hat a(\omega)\right) \label{DeltaI}\\
\hat \varphi_J &= \frac{i}{\sqrt{2}}\left(\frac{2E_C}{E_J}\right)^{\frac{1}{4}} \left(\hat \sigma^{+}-\hat \sigma^{-}\right),
\label{Phi}
\end{align}
where $\hat a$,$\hat a^{\dagger}$ and $\hat \sigma^{-}$,$\hat \sigma^{+}$ are the raising and lowering operators of the transmission line field and  the JPM states, respectively. Equations \eqref{Hamilton_Appendix}-\eqref{Phi} assuming a rotating-wave approximation, lead to the following expression for the interaction part of the Hamiltonian (infinite number of input modes):
\begin{align}
\hat H_{\rm INT} = i \hbar g \int_{-\infty}^{\infty} {\rm d}\omega (\hat a^{\dagger}(\omega) \hat \sigma^{-} - \hat \sigma^{+}(\omega) \hat a),
\end{align}
with $g \equiv (\omega_s Z_J/8\pi Z_0)^{1/2}$, where $Z_J$ is the junction impedance.

\section{Analytical solution for the continuous mean field case}
\label{app:3}

Here we give an analytical solution of the system of equations derived in Sec. \ref{sec:4A}. First we use the Laplace transformation 
$\mathcal{L}[f(t)] = f(s) = \int_0^{\infty} {\rm d}t f(t) {\rm e}^{-st}$  to rewrite the system: 

\begin{subequations}
\begin{align}
\label{Laplace_System1}
s\left<\hat \sigma^{-}(s)\right> &= - \frac{\tilde \gamma}{2} \left<\hat \sigma^{-}(s)\right>  - \rm{i} \frac{\omega_R}{2}  \left(\left<\mathcal{\hat P}_0(s)\right> - \left<\mathcal{\hat P}_1(s)\right>\right) \\ 
\label{Laplace_System2}
s\left<\hat \sigma^{+}(s)\right> &= - \frac{\tilde \gamma}{2} \left<\hat \sigma^{+}(s)\right>+ \rm{i} \frac{\omega_R}{2}  \left(\left<\mathcal{\hat P}_0(s)\right> - \left<\mathcal{\hat P}_1(s)\right>\right)  \\
\label{Laplace_System3}
s\left<\mathcal{\hat P}_0(s)\right>&= \gamma_{\rm TL} \left<\mathcal{\hat P}_1(s)\right> - \rm{i} \frac{\omega_R}{2} \left(\left<\hat \sigma^{-}(s)\right>  - \left<\hat \sigma^{+}(s)\right> \right)+1 \\
\label{Laplace_System4}
s\left<\mathcal{\hat P}_1(s)\right> &= -\tilde \gamma \left<\mathcal{\hat P}_1(s)\right> + \rm{i} \frac{\omega_R}{2} \left(\left<\hat \sigma^{-}(s)\right>  - \left<\hat \sigma^{+}(s)\right> \right)  \\
\label{Laplace_System5}
s\left<\mathcal{\hat P}_m(s)\right> &= \gamma_1 \left<\mathcal{\hat P}_1(s)\right>. 
\end{align}
\end{subequations}
The first two equations give the expressions 
\begin{align}
\left<\hat\sigma^{-}(s)\right> &= -\rm{i} \frac{\frac{\omega_R}{2}}{s+\frac{\tilde \gamma}{2}}\left(\left<\mathcal{\hat P}_0(s)\right>-\left<\mathcal{\hat P}_1(s)\right>\right) \\
\left<\hat\sigma^{+}(s)\right> &= \rm{i} \frac{\frac{\omega_R}{2}}{s+\frac{\tilde \gamma}{2}}\left(\left<\mathcal{\hat P}_0(s)\right>-\left<\mathcal{\hat P}_1(s)\right>\right),
\end{align}
which can be put into the equation for $\left<\mathcal{\hat P}_1(s)\right>$:
\begin{align}
\hspace{-0.5cm}s \left<\mathcal{\hat P}_1(s)\right> = -\tilde \gamma + \frac{\frac{\omega_R}{2}}{s+\frac{\tilde \gamma}{2}} \left(\left<\mathcal{\hat P}_0(s)\right>-\left<\mathcal{\hat P}_1(s)\right>\right).
\label{kont_Laplace_1}
\end{align}
Using the conservation of probabilities in Laplace space 
\begin{align}
\left<\mathcal{\hat P}_0(s)\right>+\left<\mathcal{\hat P}_1(s)\right>+\left<\mathcal{\hat P}_m(s)\right> = \frac{1}{s},
\end{align}
we can eliminate $\left<\mathcal{\hat P}_0(s)\right>$ in \eqref{kont_Laplace_1}:
\begin{align}
s \left<\mathcal{\hat P}_1(s)\right> = -\tilde \gamma + \frac{\frac{\omega_R}{2}}{s+\frac{\tilde \gamma}{2}} \left(\frac{1}{s}-2\left<\mathcal{\hat P}_1\right>(s)-\left<\mathcal{\hat P}_m(s)\right>\right).
\label{kont_Laplace_2}
\end{align}
Additionally, we can eliminate $\left<\mathcal{\hat P}_1(s)\right>$ in \eqref{kont_Laplace_2} with the equation for $\left<\mathcal{\hat P}_m(s)\right>$ \eqref{Laplace_System5}
\begin{align}
\hspace{-0.5cm}\left<\mathcal{\hat P}_m(s)\right> = \frac{\frac{\omega_R^2}{2}}{s\left(s+\frac{\tilde \gamma}{2}\right)\left[\frac{s^2}{\gamma_1}+\frac{\tilde \gamma s}{\gamma_1}+\frac{\frac{\omega_R^2}{2}}{s+\frac{\tilde \gamma}{2}}\left(\frac{2s}{\gamma_1}+1\right)\right]}.
\label{kont_Laplace_3}
\end{align}
To show that the numerical results of Section III give the right stationary solution, we can calculate $\lim\limits_{t \rightarrow \infty} \left<\mathcal{\hat P}_m(t)\right>$  from \eqref{kont_Laplace_3} using the relation between limits in Laplace space and real space
\begin{align}
\lim\limits_{t \rightarrow \infty} g(t) = \lim\limits_{s \rightarrow 0} s \mathcal{L}\left[g(t)\right]. \label{Laplace_limit}
\end{align}
We find
\begin{align}
\lim\limits_{t \rightarrow \infty} \left<\mathcal{\hat P}_m(t)\right> = \lim\limits_{s \rightarrow 0} s \left<\mathcal{\hat P}_m(s)\right> = 1.
\end{align}
Therefore the measurement probability in the stationary state is always one, as we have seen in the numerical results.

We next transform \eqref{kont_Laplace_3} back to real space in order to get an analytical solution for the time evolution of the measurement probability. This back transformation can be done as in Section IV using the residue theorem. The singularities of \eqref{kont_Laplace_3} are
\begin{widetext}
\begin{align*}
s_1 &= 0 \\
s_2 &= -\frac{\tilde \gamma}{2} + \frac{\tilde \gamma^2-\omega_R^2}{2\left[54(\tilde \gamma -\gamma_1)\omega_R^2+3\sqrt{36\omega_R^2\tilde\gamma^4+36(\tilde\gamma-3\gamma_1)(5\tilde\gamma-3\gamma_1)\omega_R^4+194\omega_R^6}\right]^{\frac{1}{3}}} \\
&+ \frac{\left(\frac{27\tilde\gamma\omega_R^2}{2}-\frac{27\gamma_1\omega_R^2}{2}+\sqrt{\frac{729}{4}(\gamma_1-\tilde \gamma)^2\omega_R^4+4\left(3\omega_R^2-\frac{3\tilde\gamma^2}{4}\right)^3}\right)^{\frac{1}{3}}}{3 \cdot 2^{\frac{2}{3}}} \\
s_3 &= -\frac{\tilde \gamma}{2} + \frac{\left(1+\rm{i}\sqrt{3}\right)\left(3\omega_R^2+\frac{3\tilde\gamma^3}{4}\right)}{3\cdot 2^{\frac{2}{3}}\left(-\frac{27\gamma_1\omega_R^2}{2}+\frac{27\tilde\gamma\omega_R^2}{2}+\sqrt{4\left(3\omega_R^2-\frac{3\tilde\gamma^2}{4}\right)^3+\left(-\frac{27\gamma_1\omega_R^2}{2}+\frac{27\tilde\gamma\omega_R^2}{2}\right)^2}\right)^{\frac{1}{3}}} \\
&- \frac{\left(1-\rm{i}\sqrt{3}\right)\left(-\frac{27\gamma_1\omega_R^2}{2}+\frac{27\tilde\gamma\omega_R^2}{2}+\sqrt{4\left(3\omega_R^2-\frac{3\tilde\gamma^2}{4}\right)^3+\left(-\frac{27\gamma_1\omega_R^2}{2}+\frac{27\tilde\gamma\omega_R^2}{2}\right)^2}\right)^{\frac{1}{3}}}{6\cdot 2^{\frac{1}{3}}} \\
s_4 & = s_3^{*},
\end{align*}
\end{widetext}
and the back transformation of \eqref{kont_Laplace_3} is given by
\begin{align}
\left<\mathcal{\hat P}_m(t)\right> = \frac{\omega_R^2}{2} \sum_{\stackrel{i=1}{i\neq j\neq k}}^3 \frac{\exp\left(-s_i t\right)}{\alpha_i\left(\alpha_i-\alpha_j\right)\left(\alpha_i-\alpha_k,\right)},
\end{align}
where $\alpha_i$ are the corresponding residues. Due to the first order of the singularities (all other cases are trivial), the residues are given by
\begin{align}
Res\left(s_i,\left<\mathcal{\hat P}_m(s)\right>\right) = \lim\limits_{s \rightarrow s_i} \left<\mathcal{\hat P}_m(s)\right>\left(s-s_i\right).
\end{align} 

\section{Analytical solution for the exponentially damped pulse}
\label{app:4}
In this appendix, we calculate an analytical solution for the exponentially damped pulse. We start with the Laplace transformation of the system of equations (25)
\begin{widetext}
\begin{subequations}
\begin{align}
s\left<\hat \sigma^{-}(s)\right> &= - \frac{\tilde \gamma}{2} \left<\hat \sigma^{-}(s)\right>  - \rm{i} \frac{\tilde\omega_R}{2}  \left(\left<\mathcal{\hat P}_0(s+\kappa)\right> - \left<\mathcal{\hat P}_1(s+\kappa)\right>\right) \label{Analytic_1} \\ 
s\left<\hat \sigma^{+}(s)\right> &= - \frac{\tilde \gamma}{2} \left<\hat \sigma^{+}(s)\right>+ \rm{i} \frac{\tilde\omega_R}{2}  \left(\left<\mathcal{\hat P}_0(s+\kappa)\right> - \left<\mathcal{\hat P}_1(s+\kappa)\right>\right)\label{Analytic_2}  \\
s\left<\mathcal{\hat P}_0(s)\right> &= \gamma_{\rm TL} \left<\mathcal{\hat P}_1(s)\right> - \rm{i} \frac{\tilde\omega_R}{2} \left(\left<\hat \sigma^{-}(s+\kappa)\right>  - \left<\hat \sigma^{+}(s+\kappa)\right> \right)+1\label{Analytic_3} \\
s\left<\mathcal{\hat P}_1(s)\right> &= -\tilde \gamma \left<\mathcal{\hat P}_1(s)\right> + \rm{i} \frac{\tilde\omega_R}{2} \left(\left<\hat \sigma^{-}(s+\kappa)\right>  - \left<\hat \sigma^{+}(s+\kappa)\right> \right)\label{Analytic_4}  \\
s\left<\mathcal{\hat P}_m(s)\right> &= \gamma_1 \left<\mathcal{\hat P}_1(s)\right>\label{Analytic_5},
\end{align}
\end{subequations}
\end{widetext}
with $\tilde\omega_R = \sqrt{2|\alpha|^2\kappa\gamma_{\rm TL}/\pi}$ and where we have used the relation 
\begin{align}
\mathcal{L}\left[g(t) \exp(-\kappa t)\right] = g(s+\kappa),
\end{align}
which holds for an arbitrary function $g(t)$ whose Laplace transformation exists. 

To simplify the equations we have to calculate $\left<\hat \sigma^{-}(s+\kappa)\right>$ and $\left<\hat \sigma^{+}(s+\kappa)\right>$, which can be done by multiplying \eqref{Analytic_1} and \eqref{Analytic_2} with $\exp(-\kappa t)$:
\begin{align}
\left<\hat \sigma^{-}(s+\kappa)\right> &=  \frac{-\rm{i}\frac{ \tilde\omega_R^2}{2}}{s+\kappa+\frac{\tilde \gamma}{2}} \left(\left<\mathcal{\hat P}_0(s+2\kappa)\right> - \left<\mathcal{\hat P}_1(s+2\kappa\right)\right> \label{sigma1} \\
\left<\hat \sigma^{+}(s+\kappa)\right> &=  \frac{\rm{i}\frac{\tilde \omega_R^2}{2}}{s+\kappa+\frac{\tilde \gamma}{2}} \left(\left<\mathcal{\hat P}_0(s+2\kappa)\right> - \left<\mathcal{\hat P}_1(s+2\kappa)\right>\right). \label{sigma2}
\end{align} 
Putting \eqref{sigma1} and \eqref{sigma2} into \eqref{Analytic_3} leads to 
\begin{align}
\begin{split}
 s\left<\mathcal{\hat P}_0(s)\right> &= \gamma_{\rm TL} \left<\mathcal{\hat P}_1(s)\right> +1 \\ &\hspace{-0.5cm}-  \frac{\frac{\tilde\omega_R^2}{2}}{s+\kappa+\frac{\tilde \gamma}{2}}  \left(\left<\mathcal{\hat P}_0(s+2\kappa)\right>  -  \left<\mathcal{\hat P}_1(s+2\kappa)\right>\right).
\end{split}
\end{align}
To eliminate $\left<\mathcal{\hat P}_0(s)\right>$ in this expression, we can use the conservation of probabilities in Laplace space 
\begin{align}
\left<\mathcal{\hat P}_0(s)\right> &= \frac{1}{s} - \left<\mathcal{\hat P}_1(s)\right> - \left<\mathcal{\hat P}_m(s)\right> \\
\left<\mathcal{\hat P}_0(s+2\kappa)\right> &= \frac{1}{s+2\kappa} - \left<\mathcal{\hat P}_1(s+2\kappa)\right> - \left<\mathcal{\hat P}_m(s+2\kappa)\right>,
\end{align}
which gives 
\begin{align}
\begin{split}
\hspace{-2cm} s\left(\frac{1}{s}-\left<\mathcal{\hat P}_1(s)\right> -\left<\mathcal{\hat P}_m(s)\right>\right)-1 = \gamma_{\rm TL} \left<\mathcal{\hat P}_1(s)\right> \\ -   \frac{\frac{\tilde\omega_R^2}{2}}{s+\kappa+\frac{\tilde \gamma}{2}} \left[\frac{1}{s+2\kappa}-2\left<\mathcal{\hat P}_1(s)\right>-\left<\mathcal{\hat P}_m(s)\right>\right].
\end{split}
\end{align}
Additionally, we can use \eqref{Analytic_5} to eliminate $\left<\mathcal{\hat P}_1(s)\right>$ and $\left<\mathcal{\hat P}_1(s+\kappa)\right>$:
\begin{align}
s \left<\mathcal{\hat P}_m(s)\right> &= \gamma_1 \left<\mathcal{\hat P}_1(s)\right> \\
(s+2\kappa) \left<\mathcal{\hat P}_m(s+2\kappa)\right> &= \gamma_1 \left<\mathcal{\hat P}_1(s+2\kappa)\right>.
\end{align}
Finally, we end up with the following equation:
\begin{align}
\begin{split}
\hspace{-0.5cm}\left<\mathcal{\hat P}_m(s)\right> &+ f(s) \left<\mathcal{\hat P}_m(s+2\kappa)\right> \\ &= \frac{\frac{\tilde\omega_R^2}{2}}{\left(s+2\kappa\right)\left(s+\kappa+\frac{\tilde \gamma}{2}\right)\left(s + \frac{s\left(\gamma_{\rm TL}+s\right)}{\gamma_1}\right)},\label{overall}
\end{split}
\end{align} 
with the rational function
\begin{align}
f(s) \equiv \frac{\frac{\tilde\omega_R^2}{2}\left(1+\frac{2s+4\kappa}{\gamma_1}\right)}{s+\frac{s\left(\gamma_{\rm TL}+s\right)}{\gamma_1}} \label{OM}.
\end{align} 
We are interested in the measurement probability in the stationary state, so we want to calculate $\lim\limits_{t \rightarrow \infty} \left<\mathcal{\hat P}_m\right>(t) $. To do so we use relation \eqref{Laplace_limit}.
Taking the limit on the right hand side of \eqref{OM} is straightforward, but the left hand side is more difficult. Taking a closer look at the left hand side we see
\begin{align}
\begin{split}
&\lim\limits_{t \rightarrow \infty} \mathcal{L}^{-1}\left[\left<\mathcal{\hat P}_m(s)\right> + f(s) \left<\mathcal{\hat P}_m(s+2\kappa)\right>\right] \\
= &\underbrace{\lim\limits_{t \rightarrow \infty} \mathcal{L}^{-1} \left[\left<\mathcal{\hat P}_m(s)\right>\right]}_{= \lim\limits_{t \rightarrow \infty} \left<\mathcal{\hat P}_m\right>(t)  } \\ &\hspace{-0.35cm}+ \underbrace{\lim\limits_{t \rightarrow \infty}  \int_0^t {\rm dt'} f(t') \exp\left[-2\kappa(t-t')\right] \left<\mathcal{\hat P}_m(t-t')\right>}_{\equiv (*)}.
\end{split}
\end{align} 
The first term gives us  the desired limit, while the second one describes a memory kernel that depends on the past of the system. 
To solve the integral in (*), we first have to transform $f(s)$ into real space. Since it is a rational function with only first order singularities (the other cases are trivial), $f(t)$ can be calculated using the residue theorem
 \begin{align}
 f(t') = \sum_i \alpha_i \exp\left(s_i t'\right),
 \end{align}
where $s_i$ are the singularities of the function and $\alpha_i$ the corresponding residues. The singularities are $s_1 = 0$, $s_2 = -(\kappa+\frac{\tilde \gamma}{2})$, and $s_3 = -\tilde \gamma$. Since $\left<\mathcal{\hat P}_m(t-t')\right>$ is bounded by one, the limit of the integral is determined by the exponential parts. $s_2$ and $s_3$ both damp the integrand; therefore, only the first singularity $s_1$ gives a contribution to the limit of the integral. As a result, (*) simplifies to
\begin{align}
\begin{split}
 (*) &= \lim\limits_{t \rightarrow \infty} \alpha_1 \int_0^t {\rm dt'} \exp\left[-2\kappa\left(t-t'\right)\right] \left<\mathcal{\hat P}_m(t-t')\right>\\
&\hspace{-0.4cm}\overset{(u = t-t')}{=}  \alpha_1 \int_0^{\infty} {\rm du} \exp\left[-2\kappa u\right] \left<\mathcal{\hat P}_m(u)\right>.
\end{split}
\end{align}  
If we evolve $\left<\mathcal{\hat P}_m(u)\right>$ in a Taylor expansion around zero, we can solve the integral:
\begin{align}
\begin{split}
(*) &= \alpha_1  \sum_{l=0}^{\infty} \frac{\left<\mathcal{\hat P}_m(0)^{(l)}\right>}{l!} \underbrace{\int_0^{\infty} {\rm du} \exp\left[-2\kappa u\right] u^{l}}_{= l!(2\kappa)^{-(l+1)}} \\
&= \sum_{l=0}^{\infty} \frac{\alpha_1}{(2\kappa)^{-(l+1)}} \left<\mathcal{\hat P}_m(0)^{(l)}\right>,
\end{split}
\end{align} 
where $\left<\mathcal{\hat P}_m(0)^{(l)}\right>$ denotes the $l$ th time derivative (at $t=0$). Calculating the residue
\begin{align}
\alpha_1 = \lim\limits_{s \rightarrow s_1} f(s)(s-s_1) = \frac{\tilde\omega_R^2}{2} \frac{1+4\frac{\kappa}{\gamma_1}}{\left(\kappa+\frac{\tilde \gamma}{2}\right)\left(1+\frac{\gamma_{\rm TL}}{\gamma_1}\right)}
\end{align}
and putting this all together in equation \eqref{overall}, we finally end up with an expression for the measurement probability in the stationary state:
\begin{align}
\begin{split}
\lim\limits_{t \rightarrow \infty} \left<\mathcal{\hat P}_m(t)\right> &=  \lim\limits_{s \rightarrow 0}\frac{s\frac{\tilde\omega_R^2}{2}}{\left(s+2\kappa\right)\left(s+\kappa+\frac{\tilde \gamma}{2}\right)\left(s + \frac{s\left(\gamma_{\rm TL}+s\right)}{\gamma_1}\right)} \\&\hspace{-1.5cm}- \sum_{l=0}^{\infty} \frac{\tilde\omega_R^2}{2} \frac{1+4\frac{\kappa}{\gamma_1}}{\left(\kappa+\frac{\tilde \gamma}{2}\right)\left(1+\frac{\gamma_{\rm TL}}{\gamma_1}\right)} \frac{\left<\mathcal{\hat P}_m(0)^{(l)}\right>}{(2\kappa)^{-(l+1)}}\\
&\hspace{-1.5cm}= \frac{\tilde\omega_R^2}{4\kappa\left(\kappa+\frac{\tilde \gamma}{2}\right)\left(1 + \frac{\gamma_{\rm TL}}{\gamma_1}\right)} \\&\hspace{-1.5cm}- \sum_{l=0}^{\infty} \frac{\tilde\omega_R^2}{2} \frac{1+4\frac{\kappa}{\gamma_1}}{\left(\kappa+\frac{\tilde \gamma}{2}\right)\left(1+\frac{\gamma_{\rm TL}}{\gamma_1}\right)} \frac{\left<\mathcal{\hat P}_m(0)^{(l)}\right>}{(2\kappa)^{-(l+1)}}.
\end{split}
\end{align}
The expression up to fifth order has the following form
\begin{align}
\begin{split}
\lim\limits_{t \rightarrow \infty} &\left<\mathcal{\hat P}_m(t)\right>\\ \approx &\frac{\tilde\omega_R^2}{4\kappa\left(\kappa+\frac{\tilde \gamma}{2}\right)\left(1 + \frac{\gamma_{\rm TL}}{\gamma_1}\right)} \left(1- \frac{\tilde\omega_R^2}{16\kappa^2}\right).
\end{split}
\end{align}
The validity of the approximation up to fifth order is determined by the ratio $\frac{\alpha}{\kappa}$. The smaller this ratio, the better the approximation (see Fig. \ref{fig:analytic_comparison}).

\section{Difference between photon flux and photon number}
\label{app:photon_flux}
If we take a look at the Langevin equation \eqref{Langevin} we see that the operators $\hat a_{\rm in}$ and $\hat a_{\rm in}^{\dag}$ must have unit $\sqrt{\omega}$, since $\gamma_{\rm TL}$ has units $\omega$. They cannot be unitless as the standard creation and annihilation operators. This is a side effect of input-output theory. The input and output operators are defined as a Fourier transform \eqref{a_in_def} and everything is described in terms of photon flux instead of the actual photon number. This means the value of interest is the photons arriving in a specific time interval and not the overall photon number. E.g in the continuous drive case we get
\begin{align}
\left<\hat a_{\rm in}^{\dag} \hat a_{\rm in}\right> = \frac{|\alpha|^2 \omega_0}{2\pi}.
\end{align}
We see that $\left<\hat a_{\rm in}^{\dag} \hat a_{\rm in}\right>$ describes a photon flux. The factor $2\pi$ arises from the fact that $\hat a_{\rm in}$ and $\hat a_{\rm in}^{\dag}$ are given by a Fourier transformation, which includes a respective prefactor of $1/\sqrt{2\pi}$. Here we just take care of this factor by including it into the calculations. It would also be possible to redefine $\alpha$ as $\tilde \alpha = \alpha/2\pi$, but this doesn't make a difference for the final results. Note also that $|\alpha|^2$ in this case does not describe the photon number as usual, but an amplitude of the incoming photon flux. To make a statement about the actual photon number we additionally need a time interval of interest, e.g. the measurement time $t_m$.

\section{Time dynamis of the rate equations}
\label{app:2}

Here we want to study the time evolution of the system of rate equations \eqref{rate_1}-\eqref{rate_3} for a single measurement event ($\gamma_{\rm res} = 0$). Using an algebraic computer software package, we obtain the following solution for the occupation probability of the excited state for initial conditions $P_0 = 1$ and $P_1 = 0$:
\begin{align}
P_1(t)=  K {\rm e}^{-\beta t} \sinh \left(\Gamma t\right),
 \label{Gl:A1}
\end{align}
with the constant
\begin{align}
K = \frac{8\gamma_{\rm TL}\tilde\omega}{\sqrt{\left(\gamma_{\rm TL}+\gamma_1\right)^4+16\gamma_{\rm TL}^2\left(\gamma_{\rm TL}+\gamma_1\right)\tilde \omega + 64 \gamma_{\rm TL}^2 \gamma_1^2}}
\end{align}
and the rates
\begin{align}
\Gamma &= \frac{\sqrt{16 \gamma_{\rm TL}^2 \tilde \omega  \tilde\gamma+64 \gamma_{\rm TL}^2 \tilde \omega^2+\tilde \gamma^4}}{2 \tilde \gamma} \\ 
\beta &= \frac{\tilde \gamma^2+8 \gamma_{\rm TL} \tilde \omega}{2 \tilde \gamma},
\end{align}
with $\tilde \omega \equiv |\alpha|^2\omega_0/2\pi$.
Integration of \eqref{Gl:A1} from $t'=0$ to $t'=t$ and multiplication with $\gamma_1$, together with the boundary condition $P_m(0) = 0$, lead to an expression for the measurement probability:
\begin{align}
 P_m(t) = \frac{\gamma_1 K}{\beta^2-\Gamma^2} \left[\Gamma-\Gamma \cosh(\Gamma t){\rm e}^{-\beta t}-\beta \sinh(\Gamma t){\rm e}^{-\beta t}\right]
\label{Gl:A2}.
\end{align}
In Section \ref{sec:4} we use expression \eqref{Gl:A2} to compare the rate equation approach with the mean field approach.

\end{appendix}

\bibliography{Bibliography.bib}
\bibliographystyle{apsrev4-1}

\end{document}